\documentclass[pdf]{iucr}
\journalcode{J} % Journal of Applied Crystallography

\usepackage{xcolor}
\usepackage{amsmath}
\usepackage{amssymb}
\usepackage{lineno}
\usepackage{array}

\DeclareMathOperator{\Res}{Res}

\newcommand{\un}[1]{\ensuremath{\,\mathrm{#1}}}

\begin{document}

\title{Geometrical Optics Formalism to Model Contrast in Dark-Field X-ray Microscopy}

\cauthor[a]{H. F.}{Poulsen}{hfpo@fys.dtu.dk}{}
\author[b]{L. E.}{Dresselhaus-Marais}
\author[a]{M. A.}{Carlsen}
\author[c]{C.}{Detlefs}
\author[d]{G.}{Winther}

\shortauthor{H.F. Poulsen \emph{et al.}}

\date{\today}
\maketitle

\aff[a]{Department of Physics, Technical University of Denmark, 
2800 Kgs.~Lyngby, \country{Denmark}}

\aff[b]{Lawrence Livermore National Laboratory, Physics and Life Sciences, Physics Division, 7000 East Avenue, L-487, 94550 Livermore, CA, \country{USA}}

\aff[c]{European Synchrotron Radiation Facility, 71 avenue des Martyrs, 
CS40220, 38043 Grenoble Cedex 9, \country{France}}

\aff[d]{Department of Mechanical Engineering, Technical University of Denmark, 
2800 Kgs.~Lyngby, \country{Denmark}}


\maketitle                        % DO NOT DELETE THIS LINE

\begin{abstract}

Dark-field X-ray microscopy is a new full-field imaging technique that nondestructively maps the structure and local strain inside deeply embedded crystalline elements in three dimensions. Placing an objective lens in the diffracted beam generates a magnified projection image of a local volume. We provide a general formalism based on geometrical optics for the diffraction imaging, valid for any crystallographic space group. This allows simulation of diffraction images based on micro-mechanical models.  We present example simulations with the formalism, demonstrating how it may be used to design new experiments or interpret existing ones. In particular, we show how modifications to the experimental design may tailor the reciprocal-space resolution function to map specific components of the deformation gradient tensor. The formalism supports multi-length scale experiments, as it enables DFXM to be interfaced with 3DXRD. The formalism is demonstrated by comparison to experimental images of the strain field around a straight dislocation.

\end{abstract}
\vspace{1cm}
%%%%%%%%%%%%%%%%%%%%%%%%%%%%%%%%%%%%%%%%%%%%%%%%%%%%%%%%%%%%%%%%%%%%%%%%%%%%%%

\section{Introduction}

Dark-field X-ray microscopy (DFXM) is a novel full-field imaging technique that non-destructively maps the 3D structure, orientation and strain of deeply embedded crystalline elements, such as grains or domains \cite{Simons2015, Poulsen2017, Poulsen2020}. Direct-space images are formed by placing an X-ray objective lens along the diffracted beam, affording a spatial resolution on the order of 100\un{nm}, while maintaining a working distance between the sample and X-ray objective lens that is in the cm-range.  The magnification and field of view can be adjusted by ``zooming'' in or out (a.k.a. changing the magnification), like other classical one-lens microscopes. The first implementation of a dedicated microscope was recently installed on beamline ID06 of the European Synchrotron Radiation Facility \cite{Kutsal2019}.  Since its installation, the microscope at ID06 has been used to study domain evolution in ferroelectrics \cite{Simons2018}, the austenitic transformation in shape memory alloys \cite{Bucsek2019}, recovery in metals \cite{Mavrikakis2019, Ahl2020}, embedded particles in steel \cite{Hlusko2020},  visualization of dislocation structures \cite{Jakobsen2019,Dresselhaus2020}, and the structure of biominerals \cite{Cook2018}. For related work at beamline ID01, ESRF, see \citeasnoun{Hilhorst2014} and \citeasnoun{Zhou2018}.

DFXM is conceptually similar to dark-field electron microscopy in transmission electron microscopy (TEM), which is used to selectively image strain and orientation across materials science, physics, geoscience and numerous other fields \cite{Williams1996, Nellist2000, Morones2005}.  Given its versatility, TEM based modalities based on e.g.~the insertion of apertures in the back focal plane of the objective have been adapted to DFXM to extend the use of such methods to bulk materials \cite{Poulsen2018, Jakobsen2019}.

While TEM and the analogous X-ray microscopy experiments are conceptually similar, the interaction cross sections and optical components give rise to differences between measurements with these two probes: 

\begin{itemize}
    \item \emph{Penetration.}  As DFXM uses high energy X-rays,  it can penetrate through samples that are hundreds of micrometers thick. In contrast, TEM studies are limited to thin foils with a thickness of a few hundred nanometers. DFXM can therefore be used to acquire 3D movies of the structural evolution within bulk materials while they are processed. For example, in hierarchically structured materials, it is crucial to study how the sample evolves over all relevant length scales. 
    \item \emph{Numerical aperture.} X-ray objective lenses have numerical apertures, $NA$s, that are $\sim0.001$ (covering a solid angle of $\sim{1} \un{mrad}$). The spatial resolution of DFXM is therefore fundamentally limited by the Abbe diffraction theorem to tens of nanometers. The small $NA$ of the X-ray objective also implies that it acts as a very effective filter, in both real space and reciprocal space, of stray diffraction signals, suppressing unwanted overlap and isolating an individual deeply-embedded structural element of interest.  In contrast, TEM provides atomic resolution due to a combination of a much larger $NA$ and a shorter wavelength.   
    \item \emph{Extinction length.} Strong interactions between electrons and atoms have historically limited quantitative analysis of TEM data due to the associated challenges. In contrast, high-energy X-rays have extinction lengths that are three orders of magnitude longer, making it easier to design studies that are well described by kinematical diffraction theory. This simplifies quantitative analysis for DFXM substantially.
     \item \emph{Strain mapping.} Determination of elastic strains in electron microscopy is feasible but non-trivial and subject to inaccuracies caused by a limited angular resolution and/or dynamical scattering effects \cite{Wilkinson2012}. In contrast, DFXM exhibits a high angular resolution, enabling it to map components of the strain field with high sensitivities of $10^{-4}$ or better. However, DFXM probes only one reflection meaning that not all strain components are accessible simultaneously. 
\end{itemize}

The primary differences in resolution and interaction efficiency enable DFXM to resolve a quantitative and detailed view of bulk crystals that is not attainable with DF-TEM. For this reason, the main thrust for DFXM in the future is likely to be 3D multiscale studies of microstructure and local strain evolution in hierarchically organized crystalline materials such as metals, ceramics, rocks and bones.  As it can provide high-fidelity quantitative results, DFXM can be directly compared to 3D simulations to guide, optimize and validate multiscale materials models. 

To facilitate this vision, we must develop accurate models that describe the diffraction in the sample, and the resulting imaging properties of the DFXM instrument.  \citeasnoun{Poulsen2017} used geometrical optics to derive analytical expressions for the imaging system, including the spatial resolution, field of view, aberrations and vignetting, and finished by developing a numerical model of the instrumental resolution function in reciprocal space. In \citeasnoun{Poulsen2018} a complementary presentation of the properties of the corresponding diffraction plane was derived, specifically, describing the back focal plane of the X-ray objective. However, both studies did not take crystal symmetry into consideration in their formalisms.

Another outstanding issue is in connecting DFXM results to the description of local deformation in crystals. The analysis of DFXM data so far has been based on the assumption that one can separate measurements of the local rigid body rotation and the local elastic strain. The former is probed by (large) rotation of diffraction spots at constant diffraction angle $2\theta$, the latter by (small) changes in $2\theta$ \cite{Poulsen2018}. However, micro-mechanical models and simulations typically describe deformations in terms of the deformation gradient tensor, $\mathbf{F}$, which combines rotation and strain degrees of freedom. Moreover, $\mathbf{F}$ comprises 9 independent components while a single experiment of DFXM can only probe three along a single diffraction peak. For DFXM studies to characterize individual dislocations (their Burgers vector and line direction) or other important defects or boundaries based on the limited information about $\mathbf{F}$, a more rigorous forward model is required. We currently lack a formalism for a forward model that can simulate contrast in DFXM images based on an input voxelated $\mathbf{F}$ tensor field. With such a model, the inverse problem may then be solved by iteratively comparing simulated and experimental images, while changing parameters in the materials model. (A recent demonstration of such fitting of materials parameters from 3D movies was provided by \citeasnoun{Zhang2020} for the case of 3D grain growth in Fe using DCT. More than 1000 materials parameters were determined simultaneously.)

In this paper, we revisit the diffraction imaging formalism of DFXM to address the issues mentioned above. The approach is inspired by similar work for 3DXRD, in particular work by \citeasnoun{Bernier2011}. Using the same symbols as much as possible, the presentation below allows a direct coupling between DFXM and 3DXRD algorithms. In this way, the methods may be used successively (on the same instrument) to first map all the grains in a sample with 3DXRD, then use DFXM to map a single grain with higher resolution in direct space. The formalism in this work is valid for all crystalline space groups. The equations are then given to construct a forward model of the output from continuum mechanics simulations, taking into account the highly anisotropic reciprocal-space resolution function, typically encountered with DFXM. This is supplemented by presentation of a Monte Carlo code that may be used both to sample the full form of the instrumental resolution function, $\Res$, and to construct the forward model itself.  

We also  provide a numerical example of the DFXM image that would be produced by a single $\langle 1\bar{1}0 \rangle$ edge dislocation in FCC aluminum, as derived from the long-range displacement gradient field emanating from the core. Such studies may be used both to understand contrast mechanisms, to optimize the microscope's configuration and to perform quantitative data analysis.  The result is compared to experimental data.

%%%%%%%%%%%%%%%%%%%%%%%%%%%%%%%%%%%%%%%%%%%%%%%%%%%%%%%%%%%%%%%%%%%%%%%%%%%%%%

\section{The Geometry of DFXM}

\begin{figure}
    \begin{center}
    \resizebox{0.9\columnwidth}{!}{\includegraphics{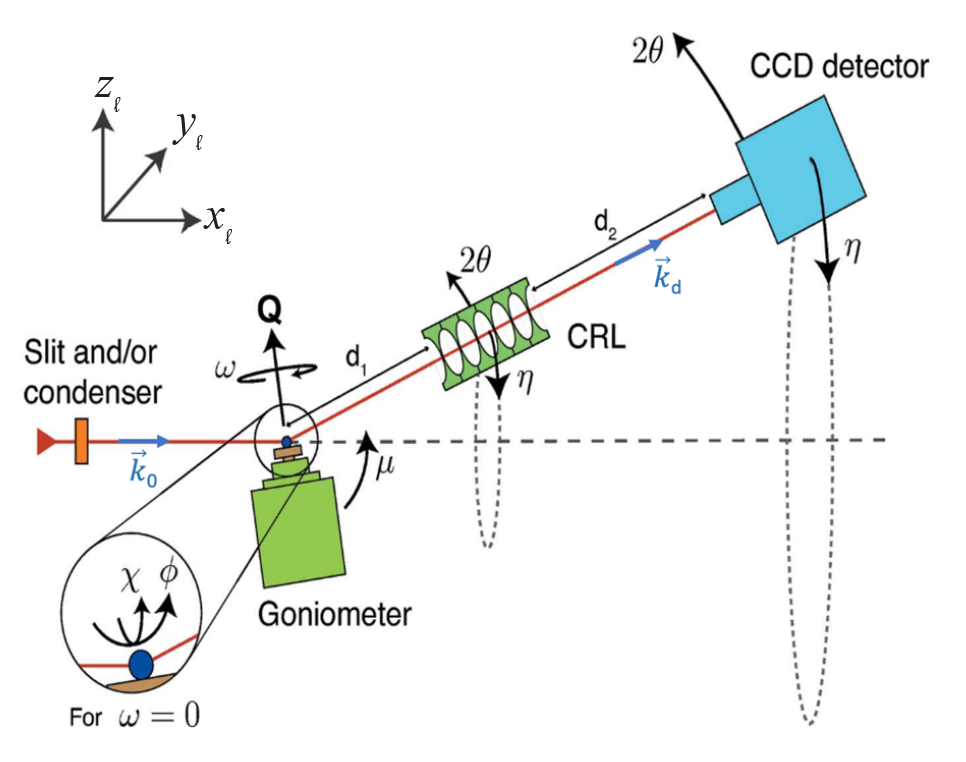}}
    \end{center}
\caption{Geometry of the dark-field X-ray microscope at ID06 at the ESRF, as shown in the laboratory coordinate system, $(x_\ell, y_\ell, z_\ell)$. We define the optical axis before the sample (shown by the red line) as the average ray of the incident X-ray beam (corresponding to the average incident wavevector $\vec{k}_{0}$, depicted as a blue arrow). We define an additional optical axis that traverses through the objective lens (CRL), which is collinear with the average diffracted ray and its corresponding wavevector $\vec{k}_d$. The pivot point of the goniometer and sample is coincident with the intersection of the two optical axes. Vector  $\vec{Q}$ defines the local scattering vector at a given point $\vec{r}(x,y,z)$ within the sample, and may be parameterized by the scattering angle $2\theta$, the azimuthal angle $\eta$ and the length of the vector $\left|\vec{Q}\right|$.  The value of $\left|\vec{Q}\right|$ is related to the spacing of the lattice plane  being measured, $d_{hk\ell}$, and the X-ray wavelength, $\lambda$, by Bragg's law. The goniometer is associated with a base tilt, $\mu$, an $\omega$-rotation around $\vec{Q}$ and two tilts, $\chi$ and $\phi$. $d_1$ is the distance from the sample to the entry point of the objective and $d_2$ is the distance from the exit point of the objective to the detector. The positive directions of the angles are indicted. This figure is adapted from \citeasnoun{Poulsen2017}.
}

\label{figure-s1}
\end{figure}

The geometry of DFXM is presented in detail in \citeasnoun{Poulsen2017}. The layout is illustrated in Fig.~\ref{figure-s1} along with a laboratory coordinate system  $(x_\ell, y_\ell, z_\ell)$. A nearly monochromatic and nearly collimated X-ray beam with average wavevector $\vec{k}_0$ illuminates the sample. This beam may be condensed in the vertical and horizontal direction  to generate a beam with a divergence that has a width of $\Delta \zeta_v$ and $\Delta \zeta_h$, respectively. The energy bandwidth is assumed to be Gaussian with a full-width half maximum (FWHM) of $\Delta E/E$. 

The goniometer is designed to access diffraction angles in a nearly vertical scattering geometry, and probe reciprocal space only in the immediate vicinity of a given reflection $(h,k,\ell)$. The current implementation of DFXM at ID06, ESRF achieves this by moving the sample along a combination of $\mu, \omega, \chi$ and $\phi$ rotation stages, see Fig.~\ref{figure-s1}.  The direction of the diffracted beam is characterized by the scattering angle, $2\theta$, and the azimuthal angle, $\eta$. We shall assume that $\eta = 0$ and $2\theta = 2\theta_0 $ for the nominal $(h,k,\ell)$ reflection. The corresponding average wavevector of the diffracted beam is $\vec{k}_d$.

The optical axis of an X-ray objective is aligned to the diffracted beam for the nominal $(h,k,\ell)$ value to produce a magnified image (inverted in both directions) on the 2D detector. The key attributes of the objective that are important to this study are the numerical aperture, $NA$, and the focal distance, $f_N$. The position and tilt of this objective defines the primary imaging system in DFXM,  which is associated with an \emph{object plane} inside the sample (at  the pivot point of the goniometer - see Fig.~\ref{figure-s2}) and an \emph{image plane} coinciding with the plane of the detector. Distance $d_1$  spans from the object plane to the entry point of the objective and $d_2$ is the distance from the exit point of the objective to the detector. The image generated by the objective has an associated magnification, $M$,  and field of view, $FOV$, and is subject to vignetting and depth of focus issues.

Early dark-field microscopy experiment used FZPs at low energies in reflection geometry \cite{Tanuma2006}. They were therefore probing only a shallow near-surface layer of the sample. For materials-science applications, higher working energies are necessary, and the objective in DFXM has typically been a Compound Refractive Lens, CRL \cite{Snigirev1996}, which is a thick lens comprising 40--80 identical parabolic lenslets \cite{Simons2017} that are made of Si, Be or a polymeric material. The resolution of DFXM in both direct and reciprocal space is presented in \cite{Poulsen2017}. Successful tests have recently been made with a pair of Multilayer Laue Lenses, MLLs, as the focusing optic \cite{Morgan2015, Murray2019}, yielding a 5--10 times larger $NA$ and a much shorter working distance and smaller $FOV$. For a detailed comparison see \citeasnoun{Kutsal2019}.

%%%%%%%%%%%%%%%%%%%%%%%%%%%%%%%%%%%%%%%%%%%%%%%%%%%%%%%%%%%%%%%%%%%%%%%%%%%%%%

\section{Formalism for Diffraction}

\subsection{Coordinate Systems} 
\label{subsec-CoordSystem}

We initially define five orthonormal coordinate systems in direct space and their corresponding collinear coordinate systems in reciprocal space. These systems describe the relationship between the  orientation of the sample, the crystallography of a grain of interest, the settings of the goniometer and objective, and the position of diffraction spots on the detector. The formalism follows \citeasnoun{Poulsen2004} and \citeasnoun{Jakobsen2019}. 

\begin{figure}
\begin{center}
\resizebox{0.9\columnwidth}{!}{\includegraphics{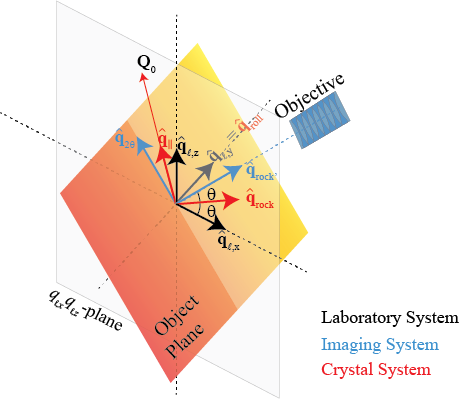}}
\end{center}
\caption{Schematic showing three of the orthonormal coordinate systems defined in Sec.~\ref{subsec-CoordSystem} in reciprocal space. These correspond to: (black)  the lab system, (red) the crystal system, and (blue) the imaging system. All vectors except for $\vec{Q}_0$ are unit vectors and all vectors except $\hat{q}_{\ell,y} = \hat{q}_{\mathrm{roll}}$ lie in the $q_{\ell,x}$-$q_{\ell,z}$ plane that is shaded translucent grey. The object plane is shown in orange and its normal vector lies along $\hat{q}_{\mathrm{rock'}}$. 
}
\label{figure-s2}
\end{figure}

The axes for the \emph{lab-based coordinate system} are defined in Fig.~\ref{figure-s1} as $(\hat{x}_{\ell}, \hat{y}_{\ell}, \hat{z}_{\ell})$. For this system, we define a collinear orthonormal reciprocal space system $(\hat{q}_{\ell,x}, \hat{q}_{\ell,y}, \hat{q}_{\ell,z})$, as depicted with black arrows in Fig.~\ref{figure-s2}.  It is natural to express the incident beam and its properties in this system.

It is natural to express the position and orientation of the objective, as well as its corresponding geometrical optics properties, in the \emph{imaging system}, which is rotated by $2\theta$ around the $\hat{y}_{\ell}$ axis with respect to the lab system. We define a collinear orthonormal reciprocal-space coordinate system, ($\hat{q}_{\mathrm{rock'}}, \hat{q}_{\mathrm{roll}}, \hat{q}_{2\theta}$). This has two coordinate axes, $\hat{q}_{\mathrm{roll}}$ and $\hat{q}_{2\theta}$, that span the object plane, as shown by the arrows in the orange object plane shown in Fig.~\ref{figure-s2}.  In this system, $\hat{q}_{\mathrm{rock'}}$ is parallel to the optical axis of the objective, which is collinear with the nominal $\vec{k}_d$. 

Next, the orthonormal \emph{crystal reference system}, shown with red arrows in Fig.~\ref{figure-s2}, is defined by $(\hat{q}_{\mathrm{rock}},  \hat{q}_{\mathrm{roll}},  \hat{q}_{\parallel})$, with $\hat{q}_{\parallel}$ parallel to the nominal $(h,k,\ell)$ lattice vector, and with $\hat{q}_{\mathrm{roll}}$ perpendicular to the vertical scattering plane (along the $\hat{q}_{\ell,x}$-$\hat{q}_{\ell,z}$ plane shown in grey in Fig. 2). The rotation axis parallel to $\hat{q}_{\mathrm{roll}}$ is the so-called ``rolling'' direction. As we did for the two other systems, we will also refer to the crystal system in terms of its collinear system in direct space.

As mentioned above, the laboratory, imaging and crystal coordinate systems all have collinear real- and reciprocal-space vectors. We shall denote position vectors in the three spaces as $\vec{r}_\ell$, $\vec{r}_i$ and $\vec{r}_c$, respectively. Likewise, we will denote diffraction vectors in the three spaces as $\vec{Q}_\ell$, $\vec{Q}_i$, and $\vec{Q}_c$, respectively.

Transformations between the three direct-space coordinate systems are given in Eq. \ref{eq-rotate_r}, and transformations between the three reciprocal-space coordinate systems are given in Eq. \ref{eq-rotate_q}, where  

\begin{align}
    \vec{r}_i & = \Theta \cdot \vec{r}_c = \Theta^2 \cdot \vec{r}_\ell \label{eq-rotate_r}
    \\
    \vec{Q}_i & = \Theta \cdot \vec{Q}_c  = \Theta^2 \cdot \vec{Q}_\ell     \label{eq-rotate_q} \\
    \Theta & = 
    \begin{pmatrix}
        \cos(\theta_0) & 0 & \sin(\theta_0) \\
        0 & 1 & 0 \\
        -\sin(\theta_0) & 0 & \cos(\theta_0)
    \end{pmatrix} .
    \label{eq1}
\end{align}

Next, we introduce a fourth orthonormal direct-space coordinate system called the \emph{sample system}, which is fixed to the sample --- again with a corresponding collinear reciprocal space system. The relationship between this system and the lab system is determined by the positions of the different goniometer stages. We identify the sample systems by the subscript $s$. Position vectors and diffraction vectors in the sample and lab systems are related by \citeaffixed{Poulsen2017}{eq.~75 in}
\begin{align}
  \vec{r}_\ell & = \mathbf{\Gamma} \cdot \vec{r}_s \\
 \vec{Q}_\ell & = \mathbf{\Gamma} \cdot \vec{Q}_s  
 \label{eq-Ql}
 \\
 \mathbf{\Gamma} & = \mathbf{M} \cdot \mathbf{\Omega} \cdot \mathbf{X} \cdot \mathbf{\Phi}.
 \label{eq-Gamma}
\end{align} 
Here $\mathbf{\Gamma}$ is the combined rotation matrix related to the $\mathbf{\Phi}(\phi)$, $\mathbf{X}(\chi)$, $\mathbf{\Omega}(\omega)$ and $\mathbf{M}(\mu)$ rotation stages of the goniometer (see Fig.~\ref{figure-s1}). Eq.~\ref{eq-Ql} represents the fundamental correlation between the measurements and the crystallography of the sample.  

Finally, we shall assume that the sample is comprised of a set of crystallographic elements such as grains, domains or subgrains. We shall assume one of these is of specific interest, and for simplicity we shall call this a ``grain''.  Hence, we define a fifth orthonormal direct space system, the \emph{grain system}, which is fixed to this selected grain of interest --- and a collinear system in reciprocal space. These are identified by subscript $g$. Positions and diffraction vectors in the \emph{sample} and \emph{grain} systems are related by
\begin{align}
    \vec{r}_s & = \mathbf{U}  \hspace{1mm}\vec{r}_g.  \label{eq-cryst2sample_direct} \\
    \vec{Q}_s & = \mathbf{U}  \hspace{1mm} \vec{Q}_g.  \label{eq-cryst2sample}
\end{align}
 As usual,  crystal symmetry must be taken into account, meaning that  \textbf{U} maps to the equivalence class 
\begin{eqnarray}
\mathbf{U}^{*} = \mathbf{U} \cdot S^{(i)} \hspace{3 mm} \forall i \in [1, N] ,
\label{eq-symmetry}
\end{eqnarray} 
where $N$ is the number of members in the relevant symmetry group $S \subset SO(3)$. Typically one operates in what is known in the literature on texture as the \emph{fundamental zone}: the member that comprises the origin of $SO(3)$ \citeaffixed{Poulsen2004}{see, e.g., p.~28 in}.

Given the small range of reciprocal space probed by DFXM, it is natural to operate with normalized diffraction vectors (in any of the systems)
\begin{eqnarray}
\vec{q} = \frac{\vec{Q} -\vec{Q}_0}{\left| \vec{Q}_0\right|}. \label{eq-qnorm}
\end{eqnarray}
These dimensionless quantities are directly related to strain components, as we shall see later.

\subsection{The Simplified Geometry}
 \label{sec-simple}
 
For simulations, it is often relevant to consider the simplest case, with a perfect alignment, where DFXM is performed in a \emph{Simplified Geometry}, with $\mu = \theta_0$, $\omega = 0$ and $ \chi = 0 $ fixed, see \citeasnoun{Poulsen2017}. Moreover we assume that during the experiment, $\phi$ is scanned (``rocked") over a small range centered around $\phi = 0$. In this case, as $\omega = 0$, the axes of rotation that define $\mu$ and $\phi$ are collinear. Consequently, we can express all coordinate transforms as simple rotations, $\mathbf{\Gamma} = \mathbf{M}(\theta_0)\cdot \mathbf{\Phi}(\phi) = \mathbf{M}(\theta_0+\phi)$. Furthermore, $\mathbf{\Theta}$ and $\mathbf{M}$ rotate about the same axis, but in opposite directions such that $\mathbf{\Theta}(\alpha)\cdot\mathbf{M}(\beta)=\mathbf{M}(-\alpha+\beta) = \mathbf{\Theta}(\alpha-\beta)$. We have

\begin{align}
  \vec{Q}_\ell = & \mathbf{M}(\theta_0+\phi)   \cdot \vec{Q}_s; \label{eq-s2l} \\
 \vec{Q}_c = & \mathbf{\Theta}(\theta_0) \cdot \mathbf{M}(\theta_0+\phi)  \cdot \vec{Q}_s = \mathbf{M}(\phi) \cdot \vec{Q}_s;\\
 \vec{Q}_i  = & \mathbf{\Theta}^2(\theta_0) \cdot \mathbf{M}(\theta_0+\phi) \cdot \vec{Q}_s = \mathbf{M}(-\theta_0 + \phi) \cdot \vec{Q}_s;   \label{eq-i2s}
\end{align}
with 
\begin{align}
     \mathbf{M} (\alpha) & = 
    \begin{pmatrix}
        \cos(\alpha) & 0 & -\sin(\alpha) \\
        0 & 1 & 0 \\
        \sin(\alpha) & 0 & \cos(\alpha)
    \end{pmatrix} .
    \label{eq-M}
\end{align}

Transformations in direct space are governed by similar expressions.

\subsection{Diffraction formalism}

From the definition of reciprocal space, the diffraction vector $\vec{Q}_g$ is related to Miller indices $(h,k,\ell)$ by  $\vec{Q}_g = \mathbf{B}\,(h,k,\ell)^T$. Following  \citeasnoun{Busing1967}, the tensor  \textbf{B} comprises the reciprocal-space lattice parameters $(a^{*}, b^{*}, c^{*},\alpha^{*}, \beta^{*}, \gamma^{*})$ as\footnote{ The choice of $\mathbf{B}$ and $\mathbf{A}$ is not unique. $\mathbf{B}$ consists of the reciprocal lattice basis $\vec{a}^\star$, $\vec{b}^\star$ and $\vec{c}^\star$ as column-vectors. These can be defined in any orthonormal coordinate system. For the analysis given here, it is essential that $\mathbf{A}^T \cdot \mathbf{B} = \mathbf{I}$. This is always the case as long as $\mathbf{A}$ is constructed, as usual, from column-vectors $\vec{a}$, $\vec{b}$, and $\vec{c
}$ with $\vec{a}^\star = (\vec{b} \times \vec{c})/V$, etc., where $V=\vec{a}\cdot(\vec{b} \times \vec{c})$.}

\begin{align}
  \mathbf{B} = 
    \begin{pmatrix}
           a^{*} & b^{*}\cos(\gamma^{*}) & c^{*}\cos(\beta^{*})\\
           0 & b^{*}\sin(\gamma^{*}) & -c^{*}\sin(\beta^{*})\cos(\alpha)\\
          0 & 0 & c^{*}\sin(\beta^{*})\sin(\alpha)
    \end{pmatrix} \label{eq-B}
\end{align}
 with 
\begin{align}
  \cos(\alpha)= \frac{\cos(\beta^{*})\cos(\gamma^{*}) - \cos(\alpha^{*})}{\sin(\beta^{*})\sin(\gamma^{*})}. \label{eq-B2}
\end{align}

In summary, this gives the result that 

\begin{align}
    \vec{Q}_s & = \mathbf{ U B} 
    \begin{pmatrix}
           h \\
           k \\
          \ell
     \end{pmatrix}. \label{eq-HFPbook}
\end{align}

For an undeformed crystal, with negligible divergence and energy spread of the incoming beam, the diffraction vector $\vec{Q}_s$ in the sample system is related to the experimental observables by \citeaffixed{Poulsen2017}{eq.~40 in} 
\begin{align}
\vec{Q}_\ell  & = \mathbf{\Gamma} \vec{Q}_s = \left| \vec{Q}\right|  
  \begin{pmatrix}
            -\sin(\theta) \\
            -\cos(\theta)\sin(\eta)\\
             \cos(\theta)\cos(\eta)
    \end{pmatrix}
    \label{eq-Qlthetaeta}
    \\
 \left| \vec{Q}\right| & =  \left| \mathbf{B} \cdot \begin{pmatrix} h \\ k \\ \ell \end{pmatrix} \right| = \frac{4\pi}{\lambda} \sin(\theta). \label{eq-Qlength}
\end{align}
Here the right hand part of the last equation represents Bragg's law.

For the simplified geometry introduced in Section \ref{sec-simple}, and with  $\Delta \theta = \theta - \theta_0$, Eqs.~\ref{eq-qnorm}, \ref{eq-s2l},  \ref{eq-Qlthetaeta} and \ref{eq-Qlength}  reduce to

\begin{equation}
    \vec{q}_s  \approx
    \begin{pmatrix}
            \phi - \Delta \theta \\
       - \cos(\theta_0) \eta \\
        \cot(\theta_0) \Delta \theta \end{pmatrix}.
        \label{eq-relief}
\end{equation}

%%%%%%%%%%%%%%%%%%%%%%%%%%%%%%%%%%%%%%%%%%%%%%%%%%%%%%%%%%%%%%%%%%%%%%%%%%%%%%

\section{Relation Between Micro-Mechanical Models and Reciprocal Space}
\label{Section-4}

In this section, we derive a revised diffraction formalism, based on the micro-mechanical concepts of the reference, undeformed sample and its corresponding crystal lattice --- which we identify with a subscript 0 --- and a deformed sample and its crystal lattice --- with no subscript. 

\subsection{Definitions from Elasticity Theory}

The deformation of a sample is typically described using elasticity theory from continuum mechanics, which we apply here to the lattice of a grain.  To express a material's deformation mechanics in reciprocal space, we begin by constructing a micro-mechanical model that is formulated in terms of the \emph{deformation gradient tensor}, $\mathbf{F}$, which we define initially in the direct-space grain system, $\mathbf{F}^g$. For a deformed specimen, $\mathbf{F}$ varies with the position in the sample, as its components are defined as $F_{ij} = \partial{x_i}/\partial{X_j}$, making it a tensor field. The transformation
\begin{eqnarray}
\vec{r}_g = \mathbf{F}^g \hspace{1mm} \vec{r}_{g,0} \label{eq-def_F}
\end{eqnarray}
relates a position in the undeformed reference system $\vec{r}_{g,0}$ to the same vector in the deformed system, $\vec{r}_g$. 

In constructing the model, we assume the direct-space lattice parameters in the undeformed state, $(a_0,b_0,c_0,\alpha_0, \beta_0,\gamma_0)$, are known but different from the analogous parameters in the deformed state, $(a,b,c,\alpha, \beta,\gamma)$. We can therefore express the deformed lattice in direct space analogously to Eqs.~\ref{eq-B} and \ref{eq-B2} by the metric 
\begin{align}
  \mathbf{A} = 
    \begin{pmatrix}
           a & b\cos(\gamma) & c\cos(\beta)\\
           0 & b\sin(\gamma) & -c\sin(\beta)\cos(\alpha^{*})\\
          0 & 0 & c\sin(\beta)\sin(\alpha^{*})
    \end{pmatrix} 
    \label{eq-A}
\end{align}
with 
\begin{align}
  \cos(\alpha^{*})= \frac{\cos(\beta)\cos(\gamma) - \cos(\alpha)}{\sin(\beta)\sin(\gamma)} .
\end{align}
Note that $\mathbf{A}^T \cdot \mathbf{B}=\mathbf{I}$, where $\mathbf{I}$ is the identity tensor and  $(\ldots)^T$ symbolises transposition.

Similarly, the undeformed state may be described by \textbf{A}$_0$, such that
\begin{align}
  \mathbf{A}= \mathbf{F}^g \mathbf{A}_0 .
\end{align}
We note that, by definition, $\mathbf{A}$, and $\mathbf{A}_0$ refer to the direct-space grain systems, whereas $\mathbf{B}$ and $\mathbf{B}_0$ refer to the reciprocal-space grain systems.

Alternatively, we can formulate the micro-mechanical model in terms of the \emph{displacement gradient tensor}, $(\mathbf{\nabla}\mathbf{u})^g$, where  $\mathbf{F}^g = (\mathbf{\nabla}\mathbf{u})^g + \mathbf{I}$. Both $\mathbf{F}^g$ and $(\mathbf{\nabla}\mathbf{u})^g$ can be written as $3 \times 3$ matrices with 9 independent components. The components of the tensor $(\mathbf{\nabla}\mathbf{u})^g$ are defined as $(\mathbf{\nabla}\mathbf{u})^g_{i,j}$, for displacement in direction $j$ of the basis vector $i$.  

Using polar decomposition, $\mathbf{F}^g$ can be split into a symmetric stretch tensor and a rotation tensor in two ways,
\begin{align}
    \mathbf{F}^g & = \mathbf{R T} = \mathbf{V R}, \label{eq-left&rightstretchtensor} 
\end{align}
where \textbf{R} is the \emph{rotation tensor}, \textbf{T} is the \emph{right stretch tensor} and \textbf{V} is the \emph{left stretch tensor} \cite{Ciarlet2004, Chadwick1999}. 

For small deformations $\mathbf{V} = \mathbf{R T R^{-1}} \approx \mathbf{T}$, and the product can be approximated as a sum, 
 \begin{equation}
    \mathbf{F}^g =  \mathbf{\epsilon}^g + \mathbf{\omega}^g + \mathbf{I},
 \end{equation}
where the symmetric tensor $\mathbf{\epsilon}^g$ is the \emph{Biot strain tensor},
\begin{equation}
    \epsilon^g_{i,j} 
    = \frac{1}{2} \left( (\mathbf{\nabla}\mathbf{u})^g_{i,j} + \mathbf{(\nabla}\mathbf{u})^g_{j,i} \right) 
    = \frac{1}{2} \left( T_{i,j} + T_{j,i} \right) - I_{ij},
    \label{eq6}
\end{equation}
where $T_{i,j}$ and $T_{j,i}$ are components of the right stretch tensor defined in Eq. \ref{eq-left&rightstretchtensor}.

From Eq.~\ref{eq6}, it follows that the strain tensor is a symmetric tensor with 6 independent components. The tensor $\mathbf{\omega}^g$ thus defines the antisymmetric part of $(\mathbf{\nabla}\mathbf{u})^g$, which corresponds to the rotation tensor in the additive decomposition of $(\mathbf{\nabla}\mathbf{u})^g$.

The tensors $\mathbf{F}^g$, $(\mathbf{\nabla u})^g$, and $\mathbf{\epsilon}^g$ are all defined in the grain system. Similar definitions apply in, e.g., the sample system. The corresponding tensors in the sample system can be found by rotating the tensors. From Eqs.~\ref{eq-def_F} and \ref{eq-cryst2sample} it follows that the deformation gradient tensor in the sample system, $\mathbf{F}^s$, is 
\begin{equation}
    \mathbf{F}^s =  \mathbf{U} \hspace{1mm} \mathbf{F}^g  \hspace{1mm} \mathbf{U}^T .
    \label{ref_rotatetensor}
\end{equation}

\subsection{Expressing Deformations in Reciprocal Space}

Following \citeasnoun{Bernier2011}, our next step is to determine the equivalent micro-mechanical model for a deformed lattice in reciprocal space.  We assume that the deformation gradient tensor in the grain system, \textbf{F}$^g$, is specified. In this case, then
\begin{align}
\mathbf{B} 
&= (\mathbf{A}^T)^{-1} 
= \big((\mathbf{F}^g \mathbf{A}_0)^T\big)^{-1}
\nonumber \\
&= \big( \mathbf{A_0}^T (\mathbf{F}^g)^T \big)^{-1}
= ((\mathbf{F}^g)^T)^{-1} (\mathbf{A}_0^T)^{-1}
\nonumber \\
&= (\mathbf{F}^g)^{-T} \mathbf{B}_0.
\label{eq-Bernier}
\end{align}
where \footnote{Note that for any orthogonal transformation $\mathbf{F}'= \mathbf{R}^T \mathbf{F} \mathbf{R}$, the inverse-transposed transforms as $(\mathbf{F}')^{-T} = \mathbf{R}^T \mathbf{F}^{-T} \mathbf{R}$.} $(\dots)^{-T} = \big((\ldots)^{T}\big)^{-1} $. In general, the deformed $\mathbf{A}$ and $\mathbf{B}$ will not have the triangular shapes of Eqs.~\ref{eq-B} and \ref{eq-A}.

Eq.~\ref{eq-HFPbook} is valid for both deformed and undeformed systems. Using \textbf{B} in this section to describe the deformed system, and   inserting Eq. \ref{eq-Bernier} in Eq.~\ref{eq-HFPbook} we get for the deformed diffraction vector
\begin{align}
  \vec{Q}_s =  \mathbf{U}(\mathbf{F}^g)^{-T} \mathbf{B}_0 
    \begin{pmatrix}
           h \\
           k \\
          \ell
     \end{pmatrix}. \label{eq-Joel}
\end{align}
As before, $\mathbf{U}$ maps to the relevant symmetry equivalence class, cf.~Eq.~\ref{eq-symmetry}.

To simplify notation, we define
\begin{eqnarray}
\mathbf{H}^g = (\mathbf{F}^g)^{-T} - \mathbf{I} \left[ \approx - ((\nabla \vec{u})^g)^T\right],  \label{eq-defH}
\end{eqnarray}
to describe our micro-mechanical model in a form that can easily be adapted to the corresponding reciprocal-space components. Inserting this in Eq.~\ref{eq-Joel}, we obtain an equation for the reciprocal lattice vectors at each position in the sample
\begin{align}
  \vec{q}_s = \mathbf{U} \hspace{1mm} \mathbf{H}^g  \hspace{1mm} \frac{\mathbf{B}_0}{ \left| \mathbf{B}_0 \cdot (h, k, \ell)^T \right|} 
    \begin{pmatrix}
           h \\
           k \\
          \ell
     \end{pmatrix}. \label{eq-Joel-norm}
\end{align}

For reference, if $\mathbf{F}$ is defined in the sample space as $\mathbf{F}^s$, we can define $\mathbf{H}^s$ similarly to Eq.~\ref{eq-defH} and subsequently rotate $\mathbf{H}^s$ into the grain system as
\begin{align}
\mathbf{H}^g  &=
    \mathbf{U}^T \mathbf{H}^s \mathbf{U}.
\end{align}

Most DFXM experiments, probe only one diffraction vector, meaning that the experiment is only sensitive to variation in the deformation fields over a 3D subspace of the 9D space spanned by all components of $\mathbf{H}$. The subspace sampled by DFXM along the ($h,k,\ell$) lattice vector is defined by the three coupled equations in Eq.~\ref{eq-Joel-norm}.  

Later we shall see that formulating the reciprocal-space model described in Eq.~\ref{eq-Joel-norm} in the imaging system can simplify the formalism for the forward model. To convert from the sample to the imaging system, we have
 \begin{eqnarray}
\vec{q}_i = \mathbf{\Theta}^2 \mathbf{\Gamma} \vec{q}_s.  \label{eq-sample2image}
\end{eqnarray}

 For a cubic crystal symmetry and the case of the grain and sample systems being identical ($\mathbf{U} = \mathbf{I}$), the expressions simplify. To describe diffraction from a $(0,0,\ell)$ reflection,
 
\begin{align}
\vec{q}_s  = & 
    \begin{pmatrix}
           H^g_{13}  \\
             H^g_{23} \\
           H^g_{33} 
     \end{pmatrix};           \label{eq-H2q}
\end{align}
 Moreover, Eq.~\ref{eq-sample2image} reduces to
\begin{align}
    \mathbf{\Theta}^2\mathbf{\Gamma}
    = & 
    \mathbf{M}(-\theta_0+\phi);
\end{align}
Hence, 
\begin{align}
\vec{q}_i  = & 
\begin{pmatrix}
    \cos(-\theta_0+\phi) H^g_{13} - \sin(-\theta_0+\phi) H^g_{33}   \\
    H^g_{23} \\
   \sin(-\theta_0+\phi) H^g_{13} + \cos(-\theta_0+\phi) H^g_{33} 
      \end{pmatrix} 
\\
        \approx & 
\begin{pmatrix}
    \cos(\theta_0) H^g_{13} + \sin(\theta_0) H^g_{33}   \\
    H^g_{23} \\
   -\sin(\theta_0) H^g_{13} + \cos(\theta_0) H^g_{33} 
      \end{pmatrix}.  \label{eq-Joel2}
\end{align}

For this simplified system, the signal that comprises images collected by DFXM include contributions that derive from only three components of the full deformation: two shear/rotational components ($H^g_{13}$ and $H^g_{23}$) and one axial ($H^g_{33}$). Previous DFXM papers have focused on this setting, with resulting mosaicity (shear) and strain scanning results \citeaffixed{Poulsen2017}{e.g.~Section 5.2 in}. 

%To build intuition for DFXM's sensitivity, it is instructive to consider a simple axial tensile deformation of $\Delta$ along the $z_g$-axis of the grain system for this simplified case. With tensile deformation, the corresponding values of $\mathbf{F}^g$ and $\vec{q}_s$ are
%\begin{eqnarray}
%\mathbf{F}^g = \begin{pmatrix}
%           1 & 0 & 0  \\
%            0 & 1 & 0  \\
%           0 & 0 & 1 + \Delta
%      \end{pmatrix} ; \hspace{2 mm}
%\vec{q}_s = \begin{pmatrix}
%           0  \\
%            0  \\
%           -\Delta
%      \end{pmatrix}.
%\end{eqnarray}
%Similarly, a simple shear deformation of $\Delta$ in the $z_g$-axis along $\hat{x}_g$ direction. Then the corresponding values of $\mathbf{F}^g$ and $\vec{q}_s$ are
%\begin{eqnarray}
%\mathbf{F}^g = \begin{pmatrix}
%           1 & 0 & 0  \\
%            0 & 1 & 0  \\
%           \Delta & 0 & 1
%      \end{pmatrix} ; \hspace{5 mm}
%\vec{q}_s = \begin{pmatrix}
%           -\Delta\\
%            0  \\
%           0
%      \end{pmatrix}.
%\end{eqnarray}
%In both cases, we see that the strain leads to a shift of the equivalent amplitude in reciprocal space and an associated sign change. 

For a given micro-mechanical model of a deformed grain, $\mathbf{A}, \mathbf{B}, \mathbf{F}^g$ and $\mathbf{H}^g$ typically vary across the grain, making them tensor fields, $\mathbf{H}^g = \mathbf{H}^g ({\vec{r}}_g)$, etc.  For each position in the grain, $\vec{r}_g$, each of these tensors may be computed explicitly using the procedure outlined above. With a known set of undeformed lattice parameters (assumed to be constant over the entire grain), one can derive $\mathbf{A}_0$, and from this $\mathbf{B}_0 = (\mathbf{A}_0)^{-T}$. These matrices may then be input into Eq.~\ref{eq-Joel-norm} or Eq.~\ref{eq-H2q} and used as a forward model to express the grain's deformation field in terms of $\vec{q}_s$. Alternatively, the model may be described using Eq.~\ref{eq-sample2image} or Eq.~\ref{eq-Joel2} to formulate the model in the imaging system, $\vec{q}_i$. In this way, we generate vector fields $\vec{q}_s(\vec{r}_s)$ and $\vec{q}_i(\vec{r}_i)$. 

%%%%%%%%%%%%%%%%%%%%%%%%%%%%%%%%%%%%%%%%%%%%%%%%%%%%%%%%%%%%%%%%%%%%%%%%%%%%%%

\section{Reciprocal-Space Resolution Function}
\label{Section-5}

The intensity acquired on the detector can be described as convolution of a model of the diffraction --- derived in Section 4 --- and an instrumental resolution function, Res. Both functions are defined in the 6D $(\vec{r}, \vec{q})$ space: the product of direct and reciprocal space. For a given detector pixel with coordinates $(y_i',z_i')$, we can write the instrumental resolution function as  $\Res(\vec{r},\vec{q},y_i',z_i')$.   
 
As the resolution function is quite complex to describe, we begin by introducing the reciprocal-space components of it for specific positions in the object plane $(y_i,z_i)$ to build intuition. First we derive the ``reciprocal-space'' resolution function for a position in the sample, $\vec{r}_0$, which lies along the optic axis (a.k.a. the ``on-axis'' position). Using a Monte Carlo approach to ray tracing simulations, we demonstrate the nature of the remaining function, which varies only in $\vec{q}$. We then extend this view to a point in the sample, $\vec{r}_1$, that lies off the optic axis (a.k.a. an ``off-axis'' position) to demonstrate how Res varies slightly over the spatial extent of the image. 

To simplify the notation, we will define the reciprocal-space version of the resolution function at $\vec{r}_0$ as $\Res_{\vec{q}}(\vec{q})$, then express it as $\Res_{\vec{q}}^{\mathrm{off}}(\vec{q}_i,y_i,z_i)$ to account for an off-axis position, $\vec{r}_1$. In all cases, the functions depend on the wavelength $\lambda$ of the X-rays and the corresponding $2\theta$ of the crystal, but not on the goniometer angles, $\mathbf{\Gamma}$ beyond its dependence on the rocking position (in the Simplified Geometry, $\phi$). We note that $\Res_{\vec{q}}(\vec{q})$ may be expressed in any of the three orthonormal coordinate systems illustrated in Fig.~\ref{figure-s2}, and point out cases for which specific coordinate systems clarify different aspects of the experiment's sensitivity.

To compute $\Res_{\vec{q}}(\vec{q})$, we first provide the relevant analytical equations to ray-trace the path of an incident ray through the microscope and onto the detector. In doing this, we constrain our ray-tracing equations to model only the rays that span the divergence and energy bandwidth of the incident beam, the rays within the angular acceptance function of the objective and the rays that can be collected by our detector, based on its specifications and the magnification. We use a Monte Carlo model to sample the intensity on the detector as a function of  $\vec{q}$, thereby deriving a numerical form for $\Res_{\vec{q}}(\vec{q})$. 

The reciprocal-space resolution function for DFXM with a CRL as the objective is presented fully in \citeasnoun{Poulsen2017} in the sample coordinate system. That work demonstrates that the response described by $\Res_{\vec{q}}(\vec{q})$ varies as a function of position across the object plane. As discussed in \citeasnoun{Poulsen2017}, the size and shape of $\Res_{\vec{q}}(\vec{q})$ can be varied dramatically with different experimental designs, for example, by  scanning $\phi$ continuously during an exposure or by inserting apertures into the diffracted beam path, e.g.~in the back focal plane of the objective. We revisit this discussion, building on it in Section 7.

\subsection{Monte Carlo Simulation of Reciprocal-Space Resolution Function}
\label{subsec-MC}

We begin by defining how a collection of rays traverse the DFXM microscope with Monte Carlo simulations. The Monte Carlo simulation procedure operates only in reciprocal space, creating an ensemble of incident rays that comply with the characteristics known for the experiment:

\begin{itemize}
    \item We define the pointing of an incident ray based on its angular offset from the optic axis, the $x_\ell$-axis, by the angles $\zeta_v$ along the vertical axis ($z_\ell$) and $\zeta_h$ along the horizontal axis ($y_\ell$). The $\zeta_v$ and $\zeta_h$ distributions are defined experimentally, typically by the divergence of the synchrotron source or condenser optics.
    
    \item The incident beam has a non-zero bandwidth, so we express the photon energy of any incident ray as its mean value $E$ plus a deviation, $\delta E/E$. The $\delta E/E$ distribution is experimentally defined. \footnote{In \citeasnoun{Poulsen2017} $\delta E/E$ was symbolized by $\epsilon$, however, we refrain from doing so here to avoid confusion with strain.}
\end{itemize}

Each ray is diffracted with equal angular probability within a solid angle defined by the objective.  We parameterize these diffracted rays with angles $\eta$ and $2\theta - 2\theta_0$, cf. Fig.~\ref{figure-s1}.  The likelihood that a diffracted ray with a given direction $(\eta,2\theta - 2\theta_0)$ is transmitted to the detector is given by the angular acceptance function of the objective. The diffracted ray has the same energy as the incident one, as the scattering is elastic. 

Bragg's law establishes a relation between the pair of incident and diffracted beams and the components of $\vec{q} = ( q_{\mathrm{rock}}, q_{\mathrm{roll}},  q_{\parallel}  ) $.  For the Simplified Geometry described above, it is shown in \cite{Poulsen2017} that the relationship can be expressed as \citeaffixed{Poulsen2017}{Eq. 55--57 in}\footnote{There is a sign error in Eq.~57 of \citeasnoun{Poulsen2017}.}
\begin{align}
    q_{\mathrm{rock}} 
    & = 
    -\frac{\zeta_v}{2} - (\theta - \theta_0) + \phi
    \label{eq2}
    \\
    q_{\mathrm{roll}}
    & = 
    -\frac{\zeta_h}{2\sin(\theta)} - \cos(\theta)\eta
 %    -\frac{\zeta_h}{2\sin(\theta)} + \frac{\eta}{2}
    \label{eq3}
    \\
    q_{\parallel}
    & =
    \frac{\delta E}{E} + \cot(\theta) \left[- \frac{\zeta_v}{2} + (\theta - \theta_0) \right].
    \label{eq4}
\end{align}
As mentioned previously, the settings of the goniometer only appear via $\phi$. Note that Eqs. \ref{eq2}, \ref{eq3}, \ref{eq4} reduce to Eq. \ref{eq-relief} for an ideal monochromatic and parallel incident beam.

The result of the Monte Carlo simulation is a 3D point cloud that describes the components of $\vec{q}$ that are sampled by a collection of incident rays that represent a given experiment. This can be transformed into a deterministic numerical model by deriving the 3D densities (binning) of a voxelized version of $\Res_{\vec{q}}(\vec{q})$. 

In principle, the Darwin width of the sample should be included in these equations, which can be done simply by adding a term to Eq. \ref{eq2}. This is, however, quite small, e.g.~for Si(111) at 14.4\un{keV}, the Darwin width is about 18\un{\mu  rad}, and is typically negligible in comparison to the vertical divergence of the incident beam. 

\subsection{Example: Top-Hat Resolution Function}
\label{subsec-MC_example}

As an example, we simulate the resolution function for a position in the sample, $\vec{r}_0=(x_i,0,0)$, that lies along the optic axis for a typical DFXM experiment. We assume that the incident beam is nearly collimated, with a divergence that is substantially smaller than the solid angle of the objective. Specifically, we simulate an experiment performed at the ESRF \cite{Dresselhaus2020,Furnace2020} with the following configuration:

\begin{itemize}
\item The incident beam was condensed only in the vertical direction, illuminating a thin layer of the sample (i.e.~a line-focused ``sheet'' beam). We assume a top-hat distribution of $\zeta_v$ values with FWHM $\Delta\zeta_v$.

\item The X-ray objective was a CRL, which  imposes a  Gaussian acceptance function upon the diffracted beam, with a FWHM that is set by the $NA$. The physical aperture of the lens, $D$, cuts off the tails of this Gaussian acceptance function, restricting resolvable diffraction angles to ones for which $2\theta-2\theta_0 < D/d_1$. This restricts the resolution function in direction $\hat{q}_{\parallel}$, as defined in Eq.~\ref{eq4}.

\item In the direction $\hat{q}_{\mathrm{roll}}$, the objective imposes a similar acceptance function upon $\xi$, where $\xi = \eta \sin(\theta)$, as appears in Eq.~\ref{eq3} (with the same Gaussian acceptance function and cut-off tails that restrict $\hat{q}_{\parallel}$, as it originates from the same optic). 

\item The energy variation of each ray, $\delta E/E$, corresponds to a distribution over all rays that may be expressed by a Gaussian with a FWHM of $\Delta E/E$. 

\end{itemize}

We implemented this function as a Monte Carlo simulation in \texttt{MATLAB} --- see \textcolor{red}{Supplementary Material}. This numerical model was run with input parameters as derived from the experiment \cite{Dresselhaus2020}: $\lambda = 0.71$\AA, $2\theta = 20.73^{\circ}$, $D=0.477\un{ mm}$ and distribution widths (all FWHM): $\Delta\zeta_v = 0.53\un{mrad}$, $NA = 0.731\times10^{-3}$, and $\Delta E/E = 1.4\times10^{-4}$. 

\begin{figure}
    \begin{center}
    \resizebox{1\columnwidth}{!}{\includegraphics{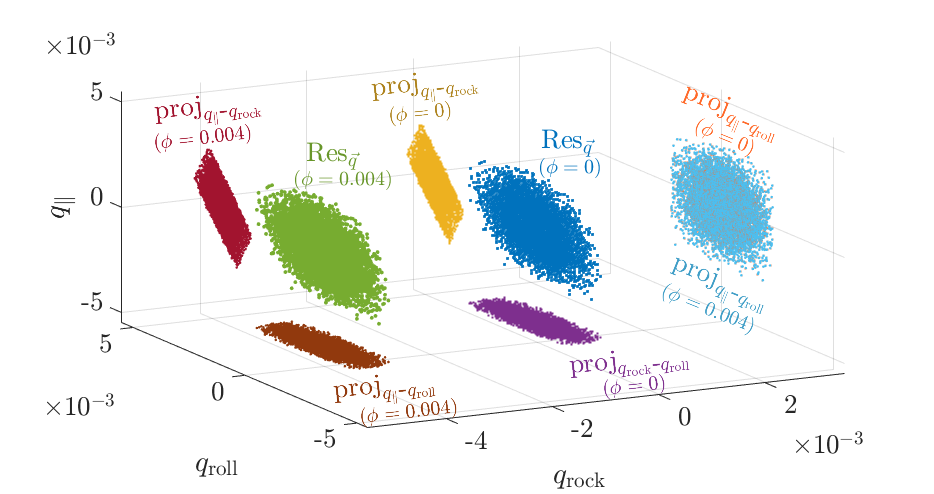}}
    \end{center}
    \caption{Numerical model of $\Res_{\vec{q}}(\vec{q})$ in the crystal system based on 10\,000 simulated rays. Dark blue cloud at center: 3D scatter plot for $\phi = 0$. The purple, orange and yellow symbols correspond to 2D projections onto the $q_{\mathrm{rock}}$-$q_{\mathrm{roll}}$ plane, $q_{\mathrm{rock}}$-$q_{\parallel}$ plane and the $q_{\mathrm{roll}}$-$q_{\parallel}$ plane, respectively. A corresponding scatter plot (green) and associated projections for $\phi = -0.004$ rad is shown to the left (in brown, red and light blue, respectively). The projection onto the $q_{\mathrm{roll}}$-$q_{\parallel}$ plane overlaps completely with that of $\phi = 0$. }
    \label{figure-s3}
\end{figure}

The plot in Fig.~\ref{figure-s3} shows the result of this numerical model in the crystal coordinate system for $\phi = 0$ and $\phi = -0.004$ rad, where each point on the scatter plot corresponds to the $\vec{q}$-components sampled by a single incident ray. 
Comparison of the projections shown in orange and light blue with those in red, yellow, brown and purple shows a large asymmetry in the reciprocal-space resolution function. To first order, the resolution function is an asymmetric box, with a ``thin direction'' parallel to the optical axis of the objective. The dimensions of the two wide axes are defined by the acceptance functions set by the numerical aperture of the objective, producing a nearly planar distribution.  The figure illustrates that the resolution function is translated along  $q_{\mathrm{rock}}$ as experiments rock the crystal during scans of $\phi$ (see Eq.~\ref{eq2}).

\begin{figure}
    \begin{center}
    \resizebox{1\columnwidth}{!}{\includegraphics{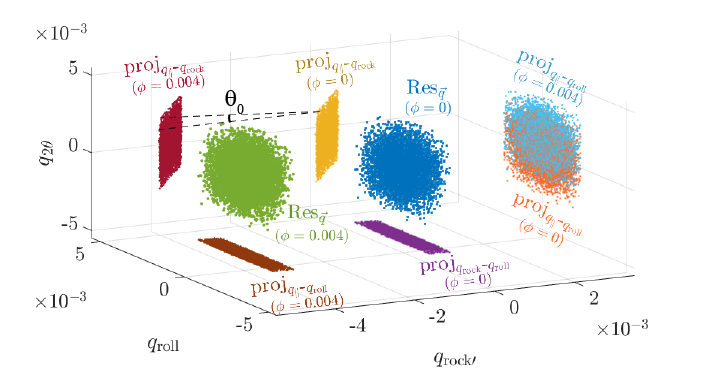}}
    \end{center}
    \caption{The same numerical model for $\Res_{\vec{q}}(\vec{q})$ as was depicted in Fig.~\ref{figure-s3}, plotted in the imaging coordinate system. The colors and symbols for this plot correspond to those of Fig.~\ref{figure-s3}. The dashed lines are guides to the eye showing that distributions are shifted in the $q_{\mathrm{rock'}}$-$q_{2\theta}$ plane when $\phi$ is changed.}
    \label{figure-s4}
\end{figure}

Figure~\ref{figure-s4} demonstrates that the main axes of the box-like distribution for $\Res_{\vec{q}}(\vec{q})$ coincide with the coordinate axes when it is plotted in the imaging coordinate system. For this reason, the thin projections along the $q_{\mathrm{rock'}}$-$q_{\mathrm{roll}}$ and $q_{2\theta}$-$q_{\mathrm{rock'}}$ planes in this system span a thinner range in the imaging system than the corresponding planes in the crystal system (by a factor of 2), suggesting that it better shows the full extent of the asymmetry in the distribution. The imaging system thus demonstrates how the angular acceptance function and divergence of the incident beam change the span of $\Res_{\vec{q}}(\vec{q})$. Fig.~\ref{figure-s4} also illustrates, however, that the shift of $\Res_{\vec{q}}(\vec{q})$ when ``rocking the sample,'' no longer aligns with the coordinate axes, as the disc-shaped distributions are shifted (without rotation) by an angle of $\theta_0$ in the $\hat{q}_{2\theta}$ direction as the sample is rotated in $\phi$.

\begin{figure}
    \begin{center}
    \resizebox{1\columnwidth}{!}{\includegraphics{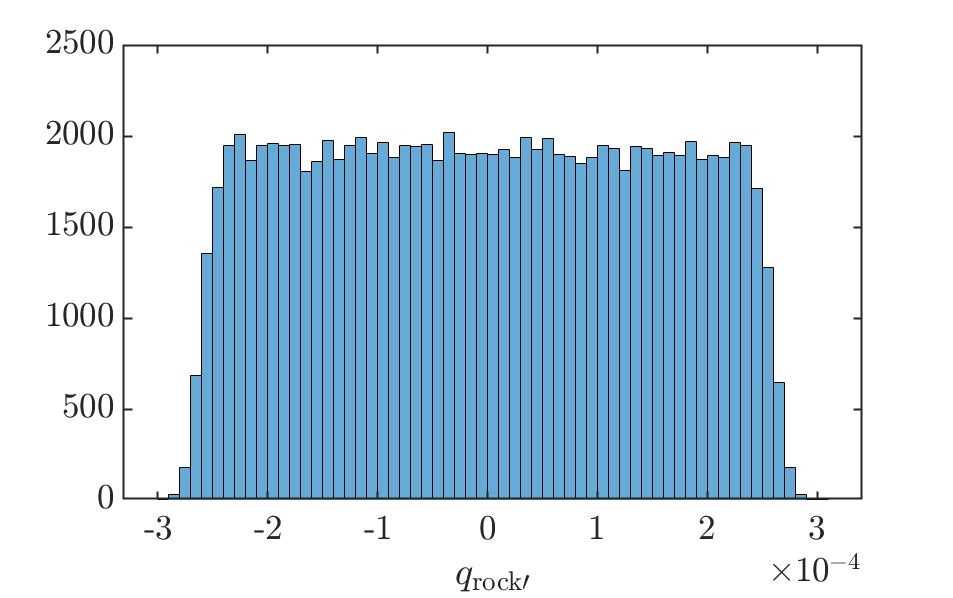}}
    \end{center}
    \caption{Width of the $q_{\mathrm{rock'}}$ distribution in Fig.~\ref{figure-s4}. To increase the accuracy, 100\,000 rays were simulated in this trace.}
    \label{figure-s5}
\end{figure}

To show the form and range of the resolution function, we show the full distribution along its narrowest axis, which in this example, aligns with the $q_{\mathrm{rock'}}$-axis, in Fig.~\ref{figure-s5} (compiled from more extensive sampling of 100\,000 simulated rays). We note that in this example that the distribution is well approximated by a top-hat function with a FWHM of 0.53\un{mrad}, as set by $\Delta \zeta_v$. In other words, for this example, the width of the narrowest portion of the resolution function (a.k.a. the strictest condition for contrast) is set most strongly by the divergence of the incoming beam. In other experimental geometries, similar simulations and projections in the imaging system could deduce the strictest condition for sensitivity by projecting the distribution along its thinnest axis.

\subsection{Generalization to Full Range of Imaging Positions}
\label{subsec-offaxis}

We now generalize this approach to positions in the object plane with $\vec{r}_1 = (x_i,y_i, z_i) \ne (x_i,0,0)$. As the center of the objective is aligned to rays traversing the central $\vec{r}_0$ portion of the sample (along the optic axis), all other portions of the object plane lie off-axis and are therefore sampled by rays that pass through the objective at an angle, shifting how the angular acceptance function interacts with the $\vec{q}$-components sampled by the X-rays. As shown in \citeasnoun{Poulsen2017}, this offset causes the reciprocal-space resolution function to shift its center, while maintaining the same shape.  The shifts in the crystal system may thus be expressed as $\Delta \vec{q}_c(\vec{r}_i)$, derived in that work as
\begin{align}
\Delta \vec{q}_c(y_i,z_i) &  = \left( {\begin{array}{cc}
   \Delta q_{\mathrm{rock}}\\ \Delta q_{\mathrm{roll}} \\ \Delta q_{\parallel}
  \end{array} } \right) = 
   \left( {\begin{array}{cc}
   \frac{\gamma z_i}{2 \tan(\theta)}\\ -\frac{\gamma y_i}{2\sin(\theta)} \\  -\frac{\gamma z_i}{2 \tan(\theta)} 
  \end{array} } \right) , \label{eq-shift1}
\end{align}
where $1/\gamma$ is a characteristic distance that may be defined for different types of objectives. For a thin lens, $1/\gamma$ is identical to the distance $d_1$ from the object plane to the plane of objective. For a thick lens, such as a CRL, $1/\gamma$ is derived in \citeasnoun{Poulsen2017}.

Rotating these expressions into the imaging system using Eq. \ref{eq-rotate_q}, we may express this as
\begin{align}
\Delta \vec{q}_i(y_i,z_i) &  = \left( {\begin{array}{cc}
   \Delta q_{\mathrm{rock'}}\\ \Delta q_{\mathrm{roll}} \\ \Delta q_{2\theta}
  \end{array} } \right) = 
   \left( {\begin{array}{cc}
   \gamma z_i \hspace{1mm} \frac{\cos(\theta)-\sin(\theta)}{2\tan(\theta)}\\ -\gamma y_i \hspace{1mm} \frac{1}{2\sin(\theta)} \\  -\gamma z_i \hspace{1mm} \frac{\cos(\theta)+\sin(\theta)}{2\tan(\theta)} 
  \end{array} } \right) . \label{eq-shift2}
\end{align}

\begin{figure}
    \begin{center}
    \resizebox{0.8\columnwidth}{!}{\includegraphics{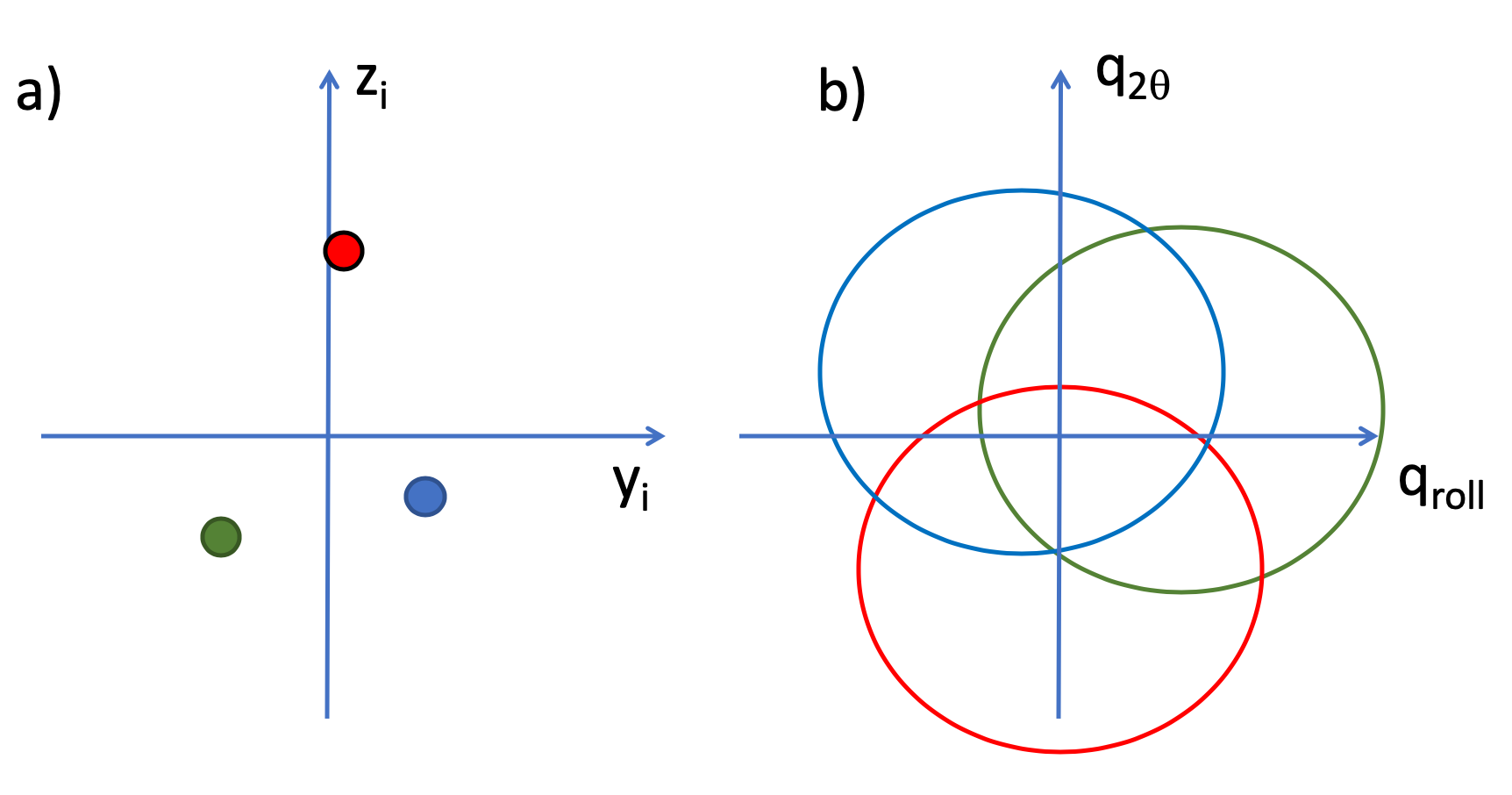}}
    \end{center}
    \caption{Coupling between position in the sample and the center of the resolution function in reciprocal space. The plot in (a) shows the $y_i$-$z_i$ plane at the object plane ($x_i=0$), specifying three random points that are distinguished by color. The plot in (b) shows the collinear reciprocal-space plane, $q_{2\theta}$-$q_{\mathrm{roll}}$, with circles corresponding to the width of the resolution function. The three circles are offset according to the $(y_i, z_i)$ positions identified by the same color.}
    \label{figure-offaxis}
\end{figure}

For a set of positions in the object plane (Fig.~\ref{figure-offaxis}a), the corresponding shifts in directions $\hat{q}_{\mathrm{roll}}$ and $\hat{q}_{2\theta}$ are illustrated in Fig.~\ref{figure-offaxis}b. The red, green and blue points in the sample and the circles outlining their corresponding $\vec{q}$-components in reciprocal space demonstrate the inversion symmetry between the collinear coordinate systems.  We can thus express the off-axis version of the resolution function in terms of the on-axis resolution function $\Res_{\vec{q}}(\vec{q})$ by
\begin{align}
\Res_{\vec{q}}^{\mathrm{off}}(\vec{q_i}, y_i,z_i) &  = \Res_{\vec{q}}(\vec{q_i}+ \Delta \vec{q}_i(y_i,z_i)).  \label{eq-offax}
\end{align}

When using a CRL as the objective, the focal length and corresponding characteristic distance $1/\gamma$ is relatively large. The example in Section \ref{subsec-MC_example} has a corresponding $1/\gamma \approx 250\un{mm}$. Then the distances of 100\un{\mu m} in the object plane are associated with shifts in $\left|\vec{q} \right|$ of $7 \times 10^{-5}$. For studies like the one presented here, the off-axis shift may be negligible, causing only an erroneous linear gradient in the simulated images. In contrast, when using MLLs as the objective, $1/\gamma$ is approximately 10\un{mm}, making these shifts important to the accuracy of the simulations.

\section{Forward Model for DFXM}

As DFXM does not measure all the tensorial components of a sample's deformation fields, a ``forward model'' can help to refine the interpretation and assign many of the lattice features that are resolved. With the formalisms for a micro-mechanical model and instrumental resolution function defined in the previous sections, we now construct the forward model to simulate how DFXM resolves a deformed crystalline sample. We shall assume that the sample only contains a single crystalline phase.

We begin this section by providing a formalism for the diffracted flux, $\Phi_d$, from a small volume in the sample and associated with a small range of reciprocal space. We then couple the diffraction model of Section~\ref{Section-4} to the instrumental resolution function described in Sec.~\ref{Section-5} to model the intensity, $I(y_i',z_i')$, that reaches a pixel of the detector when generating one DFXM image. We then explain how this model may be extended to simulate a full scan along different goniometer axes.

\subsection{Diffraction Model}

To describe the diffraction within a sample subject to the spatially-varying deformation fields described by $\mathrm{\mathbf{H}}$, we need to operate in the six-dimensional $(\vec{r}, \vec{q})$-space. The flux, $\Phi_d(\vec{r}, \vec{q})$, that diffracts from a position $\vec{r}$, with a corresponding wavevector transfer of $\vec{q}$ is
\begin{align}
    \Phi_d(\vec{r}, \vec{q}) =  C \, \Phi_0(\vec{r}) \hspace{1mm} h(\vec{r}, \vec{q}) \hspace{1mm} \rho(\vec{r})\hspace{1mm} d^3 \vec{r} \, d^3 \vec{q}, \label{eq-Id}
\end{align}
where $\Phi_0(\vec{r})$ is the incident flux at position $\vec{r}$, $\rho(\vec{r})$ is the local normalized density (a number between 0 and 1)  and $d^3 \vec{r} \, d^3 \vec{q}$ is an infinitesimal volume element in the 6D space. The pre-factor $C$ comprises scattering terms such as the Lorentz and polarization factors, and is a constant for small $\mid \! \vec{q} \! \mid$ (and therefore the same for all pixels within one DFXM image). The function $h(\vec{r}, \vec{q})$ in Eq.~\ref{eq-Id} describes the normalized likelihood of scattering in direction $\vec{q}$ at position $\vec{r}$. The scattering function is defined such that 
\begin{align}
    \iiint_{\vec{q}} h(\vec{r}, \vec{q})  \hspace{1mm} d^3 \vec{q} = 1 \hspace{3mm} \forall \hspace{2mm} \vec{r}
    %\hspace{1mm} \mathrm{when}  \hspace{1mm} \rho(\vec{r}) & = 1,
\end{align}
from which follows that
\begin{align}
    \iiint_{\vec{r}} \rho(\vec{r}) \iiint_{\vec{q}} h(\vec{r}, \vec{q})  \,\mathrm{d}^3 \vec{r} \, \mathrm{d}^3  \vec{q} & = V,
\end{align}
where $V$ is the effective volume of the diffracting object in direct space. 

Given the length-scales associated with DFXM, it is appropriate for us to assume that the deformation gradient field varies smoothly over the sample.\footnote{In crystals, discontinuities in the displacement-gradient tensor that arise from individual defects (e.g.~dislocations, stacking faults, etc.) arise from defect cores that are at least $1000{\times}$ smaller than the pixel sizes associated with DFXM. } We can thus express the deformation field at position $\vec{r}$ as $\mathbf{H}(\vec{r})$. With this notation, for each position $\vec{r}$, there is one and only one $\vec{q}_s$ (cf.~Eq.~\ref{eq-Joel-norm}  or \ref{eq-H2q}) and one and only one $\vec{q}_i$, cf.~Eqs.~\ref{eq-sample2image} or \ref{eq-Joel2}. Hence we can associate each point in direct space with one point in reciprocal space, defining the vector field $\vec{q}(\vec{r})$. Alternatively, we can use a three-dimensional Dirac delta function, $\delta^3$, to express $h$ as 

\begin{equation}
    h(\vec{q},\vec{r}) \hspace{1mm} d^3 \vec{r} \, d^3 \vec{q} = \delta^3 \big( \vec{q} - \vec{q}({\vec{r}})  \big)  \hspace{1mm} d^3 \vec{r} \, d^3 \vec{q} ,
\label{eq9}
\end{equation}

\subsection{Forward Model for a Single Image at $\phi$=0}

The result of the forward model for one image is the integrated intensity, $I$, that reaches a given pixel at the detector. As the image is a magnified and inverted version of the object plane, we can associate a position in the object plane, $(y_i, z_i)$, with a position in the image plane,  $(-My_i,-Mz_i)$, based on the magnification $M$. As already mentioned, to simplify the notation, we define coordinates $(y_i',z_i')$ to be the detector coordinates at the image plane that correspond to the analogous coordinates in the object plane ($x_i=0$).
We can then describe the integrated intensity in the pixel at  $(y_i', z_i')$ as
\begin{equation}
    I(y_i', z_i') = \iiint_{\vec{r}} \iiint_{\vec{q}}  I_d(\vec{r},\vec{q}) \Res(\vec{r}, \vec{q},y_i', z_i') \,\mathrm{d}^3 \vec{r} \,\mathrm{d}^3 \vec{q}.  \label{eq11}
\end{equation}
Here $\Res(\vec{r}, \vec{q},y_i', z_i')$ is the 6D instrumental resolution function, as introduced in Section~\ref{Section-5}. Inserting Eqs.~\ref{eq-Id} and \ref{eq9} in Eq.~\ref{eq11}, we have
\begin{align}
    I(y_i', z_i') & = C \iiint_{\vec{r}}   \Phi_0(\vec{r}) \rho(\vec{r}) \nonumber \\
    & \iiint_{\vec{q}} \Res(\vec{r}, \vec{q}, y_i', z_i') \, 
     \delta^3 \big( \vec{q} - \vec{q}({\vec{r}})  \big) \,\mathrm{d}^3 \vec{r}\,\mathrm{d}^3 \vec{q} \\ 
    & = C \iiint_{\vec{r}}   \Phi_0(\vec{r}) \rho(\vec{r}) \Res(\vec{r}, \vec{q}(\vec{r}), y_i',z_i') \,\mathrm{d}^3 \vec{r}. \label{eq11_new}
\end{align}

As shown in Eq.~\ref{eq11_new},  the intensity $I$ may be expressed as an integral over a 3D manifold in the 6D space, as $\vec{q}$ may be expressed in terms of $\vec{r}$. 
The 6D resolution function and the integrals above can be described in any of the orthonormal coordinate systems defined in Subsection~\ref{subsec-CoordSystem}. In the following, we limit the scope of our example by expressing it only in the imaging system. Equation~\ref{eq11_new} defines the fundamental ``forward model'' that describes the contrast in DFXM experimental data based on a micro-mechanical model (or any other model defining $\vec{q}(\vec{r})$) and the instrumental specifications. 

In our Monte Carlo approach, we integrate numerically by importance sampling, as is often done to define distributions related to analogous results. Specifically, the entire formalism can be programmed in terms of ray tracing.

In the following, we factor $\Res(\vec{r}_i, \vec{q}_i, y_i', z_i')$ into the direct space $\Res_{\vec{r}}$ and reciprocal space $\Res_{\vec{q}}$ resolution functions
\begin{align}
\Res(\vec{r}_i, \vec{q}_i, y_i', z_i') = \Res_{\vec{r}}(\vec{r}_i, y_i', z_i') \, \Res_{\vec{q}}(\vec{q}_i, y_i', z_i'). \label{eq-r&q}
\end{align}

The direct-space component of the resolution function, $\Res_{\vec{r}}(\vec{r}_i,y_i',z_i')$, spans a finite gauge volume that is defined in $(y_i,z_i)$ by the intrinsic spatial resolution related to the objective and by the point-spread function (PSF) of the detector. Assuming geometrical optics and kinematical diffraction, an expression for the intrinsic spatial resolution is provided in \citeasnoun{Poulsen2017}. For a typical detector with a scintillator coupled to the camera by visible light microscopy optics, the PSF varies isotropically over $y_i'$ and $z_i'$. 
The resolution in direction $x_i$ is given by the depth of field of the objective. In the following we shall for simplicity of presentation neglect both the intrinsic spatial resolution and the depth of focus issue, but the formalism is easily extended to include these effects. Then
\begin{align}
    \Res_{\vec{r}}(\vec{r_i}, y_i', z_i')  & = \mathrm{PSF}(y_i - y_i',z_i - z_i') , \label{eq_res_r}
\end{align}
where $\mathrm{PSF}$ is the 2D point spread function. We note that the missing $x_i$ coordinate on the right hand side of Eq.~\ref{eq_res_r} implies that the resolution function imposes a 1D integration of the deformation field in the sample along a direction normal to the object plane. 

The reciprocal-space resolution function was derived for positions in the object plane that are off-axis  in Eq.~\ref{eq-offax}. Inserting this and Eqs.~\ref{eq-r&q} and \ref{eq_res_r} into Eq.~\ref{eq11_new}
\begin{align}
    I(y_i', z_i') & = C \int_{y_i}  \int_{z_i} \mathrm{PSF}(y_i - y_i', z_i - z_i') \nonumber\\
    &\int_{x_i}  \Phi_0(\vec{r}_i) \rho(\vec{r}_i) \Res_{\vec{q}}(\vec{q_i}(\vec{r}_i)+ \Delta \vec{q}_i(y_i',z_i')) \, \mathrm{d}y_i \mathrm{d}z_i \mathrm{d}x_i. \label{eq-forward2}
\end{align}

The reciprocal-space resolution function and the point-spread functions may be approximated with analytical functions or look-up-tables to a good accuracy. The forward model can then be generated by numerical integration, which is faster than ray tracing. 

Returning to the reciprocal-space imaging system, the disc-shaped distribution that was introduced in Section \ref{subsec-MC_example}) can also enable the full resolution function to be factored to a good approximation
\begin{align}
  \Res_{\vec{q}}(\vec{q_i}(\vec{r}_i)) = \Res_{\mathrm{rock}'}(q_{\mathrm{rock'}}(\vec{r}_i))
  \Res_{\mathrm{roll}}(q_{\mathrm{roll}}(\vec{r}_i))
  \Res_{\mathrm{2\theta}}(q_{\mathrm{2\theta}}(\vec{r}_i)).
  \label{eq-q_split}
\end{align}

As was shown before, for the full $\Res(\vec{r},\vec{q}(\vec{r}),y_i',z_i')$, the imaging system simplifies comparison between the $NA$ of the objective and the extent of the instrumental resolution function. In cases where the $NA$ is sufficiently large that it extends beyond most variation in $q_{\mathrm{roll}}$ and $q_{\mathrm{2\theta}}$ based on its micro-mechanical model, the variations in these directions are effectively ``integrated out.'' This means that for all positions, $\vec{r}$, we have $Res_{\mathrm{roll}}(q_{\mathrm{roll}}(\vec{r})) =  \Res_{\mathrm{2\theta}}(q_{\mathrm{2\theta}}(\vec{r})) = 1$. Then Eq.~\ref{eq-forward2} reduces to 
\begin{align}
    I(y_i', z_i') & = C \int_{y_i} \int_{z_i} \mathrm{PSF}(y_i - y_i', z_i - z_i') \nonumber\\
    &\int_{x_i}  \Phi_0(\vec{r}_i) \rho(\vec{r}_i) \Res_{\mathrm{rock}'}(q_{\mathrm{rock'}}(\vec{r}_i))+ \Delta q_{i,1}(y_i',z_i')) \, \mathrm{d}y_i \mathrm{d}z_i \mathrm{d}x_i. \label{eq-forward3}
\end{align}

Next, if one neglects the finite width of the PSF and assumes a uniform illumination with no attenuation, the expression reduces to a 1D integral
\begin{align}
    I(y_i', z_i') & = C^* \int 
    \Res_{\mathrm{rock}'}(q_{\mathrm{rock'}}(x_i,y_i',z_i')+ \Delta q_{i,1}(y_i',z_i')) \, \mathrm{d}x_i, \label{eq-forward4}
\end{align}
where $C^*$ is the adjusted constant that includes the extra terms integrated out of Eq.~\ref{eq-forward3}. Moreover, if $ \Res_{\mathrm{rock}'}$ is a top-hat function like the case presented in Fig.~\ref{figure-s5}, the forward model reduces to the length of line segments spanning the $\vec{q}$-range defined by the top-hat function.  This was the approach of the first DFXM forward model of dislocations, see \citeasnoun{Jakobsen2019}.

\subsection{Forward Model for a Scan}
\label{section-forward&scan}

During a DFXM experiment, a stack of images are typically acquired over a series of $(\mu, \chi, \phi, \omega, 2\theta, \eta)$ settings. The formalism of the preceding section can be used to simulate the corresponding series of images for a given scan, provided the following changes are made
\begin{itemize}
    \item \emph{Revision of the relation $\vec{q} = \vec{q}(\vec{r})$:} The motor movements will change the relationship between $\vec{q}_i$ and $\vec{q}_s$, cf. Eq. \ref{eq-sample2image}, or for the simplified setting Eq. \ref{eq-i2s}. This implies that the sample rotations correspond to rotation of the elements of \textbf{H} that contribute to the signals measured in the virtual experiment.
    \item \emph{Shifts of resolution element:} In the Simplified Geometry, when $\phi \neq 0$ the reciprocal space resolution element is shifted in the $q_{\mathrm{rock'}}$-$q_{2\theta}$ plane, cf. Fig. \ref{figure-s4}. Likewise other motor movements offsets the resolution element.
\end{itemize}

\section{Designing Experiments to Sample Components of the Deformation Gradient Tensor}

In this section, we discuss how new insights from our forward model can  guide experimental design and data collection. Our primary aim is to minimize the data acquisition time necessary to collect fully interpretable datasets that can enable DFXM to study material dynamics.  

We note that Res$_{\vec{q}}(\vec{q}$) can be modified in many ways by inserting apertures in the back-focal plane \cite{Jakobsen2019}. Very localized features in such (coded) apertures will, however, diminish the spatial resolution, cf. the Abbe diffraction limit \cite{Born2013}. Likewise the width of the resolution function in direction $\hat{q}_{rock}$ can be made arbitrarily large by rotating the sample in $\phi$ during an exposure \cite{Poulsen2017}, while the width in the $\hat{q}_{roll}$ direction can be made arbitrarily large by rotating the sample in $\chi$ during an exposure.  Changing the shape of the resolution function can be relevant for faster characterization of larger regions in reciprocal space. It may also improve the accuracy when sampling larger volumes in reciprocal space by combining volume elements by avoiding overlap or voids between these volume elements. These aspects are discussed in \citeasnoun{Poulsen2017}.  

One very fast approach to data acquisition is based on inserting a circular beamstop in the back focal plane to block the central part of the beam. With this approach, the signal from an undeformed region of the material can be separated from the signal of the deformed regions surrounding defects. With a beamstop in the back focal plane, integration over $\phi$ during an exposure can allow a single image to visualize the regions in the grain that are mostly pristine or the regions that are highly deformed or strained, without needing to resolve the specific components of \textbf{H}$^g$ that contribute to the signal. We note that this approach provides the opportunity to optimize the inherent \textbf{H}-resolution of the instrument during experiments on samples with deformations that are not as well known. For these experiments, initial scans could qualitatively use the back-focal-plane beamstop for contrast, then integrate over successively smaller rotations of $\phi$ to determine the extent of the deformation fields.

Another opportunity is to use the forward model in this work to optimize the design of the experiment, such that the sensitivity of one image effectively integrates $\Res_{\vec{q}}(\vec{q})$ over a 2D plane of the 3D orthogonal reciprocal space. By translating the position of such a plane --- facilitated by scanning one or more of the degrees of freedom of the instrument --- one can then effectively probe one component of \textbf{H}$^g$. Below we present an example of such an approach.

\subsection{Mapping Three Components of \textbf{H}$^g$ using a Top-hat Resolution Function}
\label{sec-platelike}

Inspired by Figs.~\ref{figure-s3} and \ref{figure-s4}, the aim of this approach is to design a reciprocal-space resolution function that extends so far in both the $q_{\mathrm{roll}}$ and $q_{2\theta}$ directions that any variation in these directions are ``integrated out''. The instrument is then sensitive to only $q_{\mathrm{rock}'}$  and the resolution function becomes one-dimensional in reciprocal space.

\begin{figure}
    \begin{center}
    \resizebox{1.0\columnwidth}{!}{\includegraphics{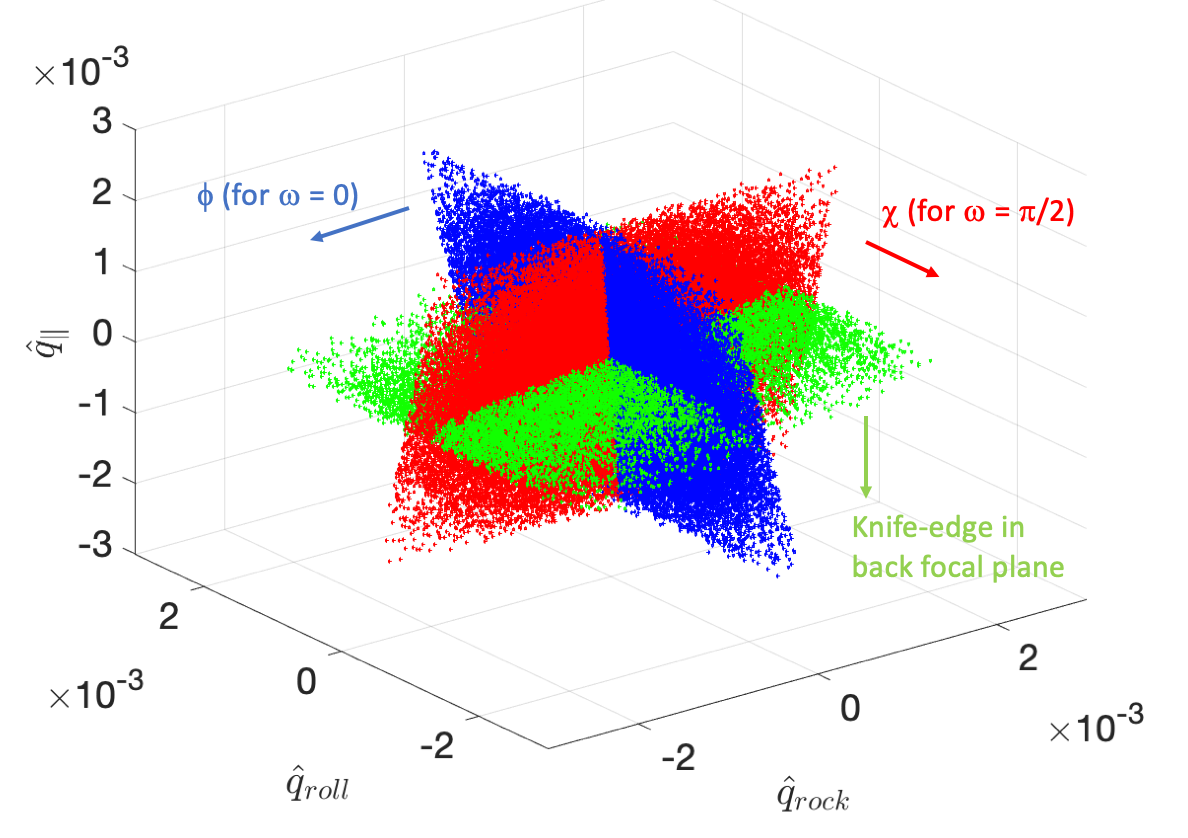}}
    \end{center}
    \caption{Illustration of sampling of components of \textbf{H}$^g$ using three almost orthogonal scans in orthonormal crystal space. The reciprocal space resolution function are shown as a scatter plot for the three settings (red, green and blue).
    The relevant goniometer motors to scan are indicated. The instrumental settings are similar to the simulations of Section \ref{subsec-MC_example}. For the green setting the image was integrated over a range of 0.004\un{rad} in $\phi$.  }
    \label{figure-scansin3D}
\end{figure}

For the special case of  a  cubic  crystal  symmetry and  diffraction from a $(00\ell)$ reflection, the contrast is then defined simply by the first component of $\vec{q}$ in the imaging system. As already derived in Eq. \ref{eq-Joel2},
\begin{equation}
    q_{i,1} (\omega = 0) = \cos(\theta_0)H^g_{13} + \sin(\theta_0)H^g_{33}. \label{eq-hallelujah}
\end{equation}

If the width of Res$_{\mathrm{rock}'}$ is sufficiently small, it may be neglected, removing the need for deconvolution in reciprocal space. Acquiring images while ``rocking'' (scanning $\phi$) then generates a distribution comprised of a linear combination of two components of $\mathbf{H}^g$, as identified in Eq.~\ref{eq-hallelujah}, for each pixel (and corresponding volume $V$ in the sample). We illustrate this condition with the blue points showing the components probed in Fig.~\ref{figure-scansin3D}.

It appears that the ideal shape of $\Res_{\vec{q}}(\vec{q})$ in this case is that of a thin rectangular plate. This can be obtained by maximizing the $NA$ of the objective and the horizontal divergence while minimizing the vertical divergence and energy bandwidth. In the case of a parallel incoming beam, the width (FWHM) in the $\hat{q}_{\mathrm{rock}'}$ direction becomes $(\Delta E/E) \sin(\theta)$, cf.~Eq.~\ref{eq4}.  For the example in Section \ref{subsec-MC_example} this corresponds to $2.6\times 10^{-5}$. 

Rotating the sample by $90^{\circ}$ in $\omega$ also rotates our resolution of $\mathbf{H}^g$, sampling a component that is nearly orthogonal to the previous one, with a similar shape. 
\begin{eqnarray}
q_{i,1}(\omega = \pi/2) =  -\cos(\theta_0)H^g_{23} + \sin(\theta_0)H^g_{33}.
\label{eq-amen}
\end{eqnarray}
In this configuration, acquiring a series of images while scanning over $\chi$ associates each pixel on the detector (and corresponding gauge volume, $V$, in the sample) with a distribution of this second component of $\mathbf{H}^g$. This configuration is illustrated by the red points in Fig.~\ref{figure-scansin3D}.

To measure the third nearly-orthogonal component, we propose to insert a knife-edge in the back-focal plane (e.g.~at $\omega = 0$).  If the blade of the knife is horizontal in the direct-space system, it truncates $\Res_{\vec{q}}(\vec{q})$ in the vertical direction. Acquiring a series of images that scan this knife-edge, one may deconvolve the effect of the knife-edge and associate each pixel on the detector (and corresponding volume $V$ in the sample) with a distribution of this third combination of components. To integrate over two directions in reciprocal space, we  scan $\phi$ over a certain range while acquiring data. The scanning direction of the knife-edge is parallel to $\hat{q}_{2\theta}$; as a result this becomes a scan of  
\begin{eqnarray}
q_{i,3}(\omega = 0) =  \sin(\theta_0)H^g_{13} + \cos(\theta_0)H^g_{33}.
\label{eq-dominietsanctus}
\end{eqnarray}
This scanning configuration is illustrated by the green points in Fig.~\ref{figure-scansin3D}. The conversion factor between the translation of the knife edge and step in $q$-space is provided in \citeasnoun{Poulsen2018}. (The insertion of a narrow slit in the back focal plane will compromise spatial resolution, see \citeasnoun{Poulsen2018}.)

%%%%%%%%%%%%%%%%%%%%%%%%%%%%%%%%%%%%%%%%%%%%%%%f%%%%%%%%%%%%%%%%%%%%%%%%%%%%%%%

\section{Example: Forward Modelling the Deformation Field around a Pure Edge Dislocation}

To demonstrate the validity of our forward model, we include an example that describes a simple system for which we have experimental data. In this case, we model a single, isolated, edge dislocation in single-crystal aluminum in its FCC structure. The results of this model may be compared directly to the full interpretations presented in \citeasnoun{Dresselhaus2020}, where the full analysis of the dislocation mechanics is discussed explicitly. To compile this model using Eq.~\ref{eq11}, we first define how analytical expressions of the mechanics are defined in the diffracted beam, then we describe the instrumental resolution function. Finally, we detail how the simulations were carried out and provide a comparison with experimental data.

\subsection{The Displacement Gradient Field of an Edge Dislocation}
\label{section-displacement_ballroom}

\begin{figure}
    \begin{center}
    \resizebox{1\columnwidth}{!}{\includegraphics{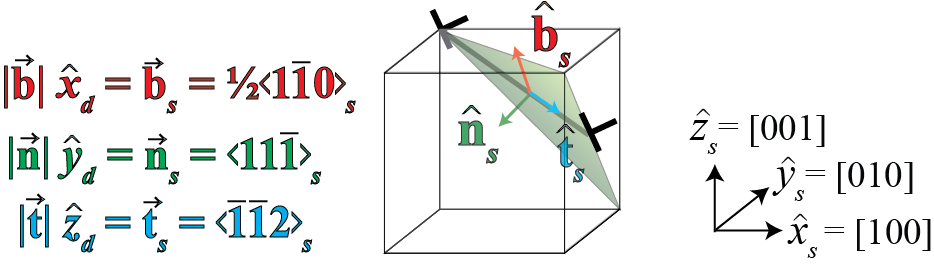}}
    \end{center}
\caption{A straight edge dislocation defined by Burgers vector $\vec{b}$, line direction $\vec{t}$ and glide plane normal $\vec{n}$ in an FCC crystal lattice and its relation to the present sample/grain system. See text. }%THE INDICES OF b, t AND n DO NOT MATCH. SUGGEST TO PUT THEM ALL IN <> INSTEAD OF [] FOR NOW}
\label{figure-dislocationSystem}
\end{figure}

We begin by defining a micro-mechanical model for the distortions surrounding a pure, straight edge dislocation. The dislocation is defined by its Burgers vector, $\vec{b}$, its line direction, $\vec{t}$, and the normal to its glide plane, $\vec{n}$. In the FCC lattice, these directions align with $\{ 110 \}$,  $\{ 112 \}$ and  $\{111\}$, respectively. This  is illustrated for one of the 12 symmetry-related systems in Fig.~\ref{figure-dislocationSystem}. We define an orthonormal \emph{dislocation coordinate system} by $(\hat{b}, \hat{n}, \hat{t})$, as shown in Fig.~\ref{figure-dislocationSystem}.

As this experiment studied a single crystal, the sample and grain systems are identical in this case, meaning that \textbf{U} = \textbf{I}. To simplify the coordinate transforms between the sample and dislocation systems, we define the lattice parameters to lie along the coordinate axes of the sample system. For this reason, the rotation matrix, $\mathbf{R}^{s,d}$, that transforms a vector from the sample system into the dislocation system is defined with
\begin{align}
\vec{r}_d & = \mathbf{R}^{s,d} \vec{r}_s \\
\nonumber
\mathbf{R}^{s,d} & = \begin{pmatrix}
           b_{x,s} & b_{y,s} & b_{z,s}  \\
           n_{x,s} & n_{y,s} & n_{z,s}  \\
           t_{x,s} & t_{y,s} & t_{z,s}
      \end{pmatrix}, \hspace{5 mm}
\end{align}
where $\mathbf{R}^{s,d}$ is comprised of the $x,y,z$ components of the unit vectors $\hat{b}$, $\hat{n}$ and $\hat{t}$ in the sample frame.

The displacement field, $\vec{u}_{d}= (u_{d,x},u_{d,y}, u_{d,z})$, around the dislocation is defined by \citeasnoun{Hirth1992} as
\begin{align}
u_{d,x} & = \frac{b}{2\pi}\Big[\tan^{-1} \!\! \Big(\frac{y_d}{x_d} \Big) + \frac{x_dy_d}{2(1-\nu)(x_d^2 + y_d^2)} \Big]   \label{eq-Lothe1}, \\
u_{d,y} & = - \frac{b}{2\pi} \Big[ \frac{1-2\nu}{4(1-\nu)}  \ln(x_d^2 + y_d^2) + \frac{x_d^2-y_d^2}{4(1-\nu)(x_d^2 + y_d^2)}  ) \Big]   \label{eq-Lothe2}, \\
u_{d,z} & = 0,  
\label{eq-Lothe3}
\end{align}
where $b$ is the length of the Burgers vector and $\nu$ is the Poisson ratio. In the present example of aluminium at elevated temperature,  $b = 3.507$ {\AA} and $\nu = 0.334$.

The deformation gradient tensor in the dislocation system, $\mathbf{F}^{d}$, is by definition 
\begin{equation}
   \mathbf{F}^{d} =   \mathbf{I}  +
\begin{pmatrix}
\frac{\partial u_{d,x}}{\partial x} &  \frac{\partial u_{d,x}}{\partial y} & 0 \\
\frac{\partial u_{d,y}}{\partial x} & \frac{\partial u_{d,y}}{\partial y} & 0 \\
0 & 0 & 0
\end{pmatrix},
\end{equation}
where the non-zero components are
\begin{align}
F^{d}_{xx} & =  1 -\frac{by_d}{4\pi(1-\nu)} \Big[\frac{3x_d^2 +y_d^2 - 2\nu(x_d^2+y_d^2)}{(x_d^2+y_d^2)^2} \Big], \\
F^{d}_{xy}  & =   \frac{bx_d}{4\pi(1-\nu)} \Big[\frac{3x_d^2 +y_d^2 - 2\nu(x_d^2+y_d^2)}{(x_d^2+y_d^2)^2} \Big], \\
F^{d}_{yx} &=  -\frac{bx_d}{4\pi(1-\nu)} \Big[\frac{x_d^2 + 3y_d^2 - 2\nu(x_d^2+y_d^2)}{(x_d^2+y_d^2)^2} \Big],\\
F^{d}_{yy} & = 1 +\frac{by_d}{4\pi(1-\nu)} \Big[\frac{x_d^2 - y_d^2 - 2\nu(x_d^2+y_d^2)}{(x_d^2+y_d^2)^2} \Big],\\
F^{d}_{zz} & = 1.
\end{align}

$\mathbf{F}^d$ may be rotated by $\mathbf{R}^{s,d}$ into the sample space to calculate $\mathbf{H}^s = \mathbf{H}^g$
\begin{align}
   \mathbf{H}^s  & = {( (\mathbf{R}^{s,d}})^T \mathbf{F}^d {\mathbf{R}^{s,d}})^{-T} - \mathbf{I},   \nonumber\\
& = ({\mathbf{R}^{s,d}})^T(\mathbf{F}^d)^{-T} {\mathbf{R}^{s,d}} - \mathbf{I}.  \label{eq-defH_disloc}
\end{align}
For the example used below, we define the dislocation line to pass through the origin of both the coordinate systems.\footnote{We note that if we had not defined our coordinate system as such, coordinate transforms from the dislocation system to the sample system would require both the rotation matrix provided and a translation matrix.} 

\subsection{Geometry and Resolution Function}

Following the experiments by \citeasnoun{Dresselhaus2020}, we shall use the Simplified Geometry, with a goniometer set to $\mu = \theta$, $\omega = 0$, $\chi =0 $, with a single crystal $\mathbf{U} = \mathbf{I}$, and diffraction from a (002) reflection. The cubic crystal symmetry then implies that $\vec{q}_i$ is given by the simple relation in Eq. \ref{eq-Joel2}. 

In this experiment, the sample was illuminated with a line beam that had a spatial intensity profile that we assume to be Gaussian along $z_{\ell}$ and homogeneous in $y_{\ell}$ across our field of view, where the FWHM of $\Delta z_{\ell}  = 0.6 \mu$m. The sample is assumed sufficiently large that the volume inspected is defined by the size of the illuminating beam and the field-of-view of the camera, corresponding to a 30 x 30 $\mu$m$^2$ region within the sample in the object plane, as defined in the imaging system. The effective pixel size in the object plane is $p =$ 75 nm, and the density of the material is constant, so $\rho(\vec{r}) = 1$ throughout.  The general details of the experimental setup, including details of the objective, were described in Section \ref{subsec-MC_example}. 

The shape of the reciprocal-space resolution function, Res$_{\vec{q}}(\vec{q})$, for this set-up were presented in Figs \ref{figure-s3} and  \ref{figure-s4}. The $NA$ of $0.73\times10^{-3}$ (FWHM) implies that the resolution function has a half width at half maximum (HWHM) in the rolling direction of $NA/(4\sin(\theta))$ = 0.0011. Likewise, the resolution function has a HWHM in the $2\theta$ direction of $NA/(4\tan(\theta))$ = 0.0011. Hence, components of \textbf{H}$^g$ are approximately integrated when they numerically smaller than  0.0011.  In comparison, the displacement-gradient field from the edge dislocation decays as $b(3-2\nu)/(4\pi(1-\nu)r) = 0.078/r$ , where $r$ is the radial distance away from the nearest point along the dislocation line, as measured in nm. It appears that the ``integrated'' form of $\Res_{\vec{q}}(\vec{q})$ is a reasonable approximation for $r > 70$ nm. Given the spatial resolution of the DFXM instrument, this implies that the integration is a reasonable approximation for the forward model except near the core. Nevertheless, to provide as accurate a simulation as possible, Res$_{\vec{q}}(\vec{q})$ was simulated with the Monte Carlo code presented in section \ref{subsec-MC}, and the result was transformed into a voxelized density function. Moreover, as the characteristic distance $1/\gamma$ is of order 25 cm, we will neglect the off-axis shifts detailed in Eq. \ref{eq-shift2} and Fig.\ref{figure-offaxis}.

\begin{figure}
    \begin{center}
    \resizebox{1\columnwidth}{!}{\includegraphics{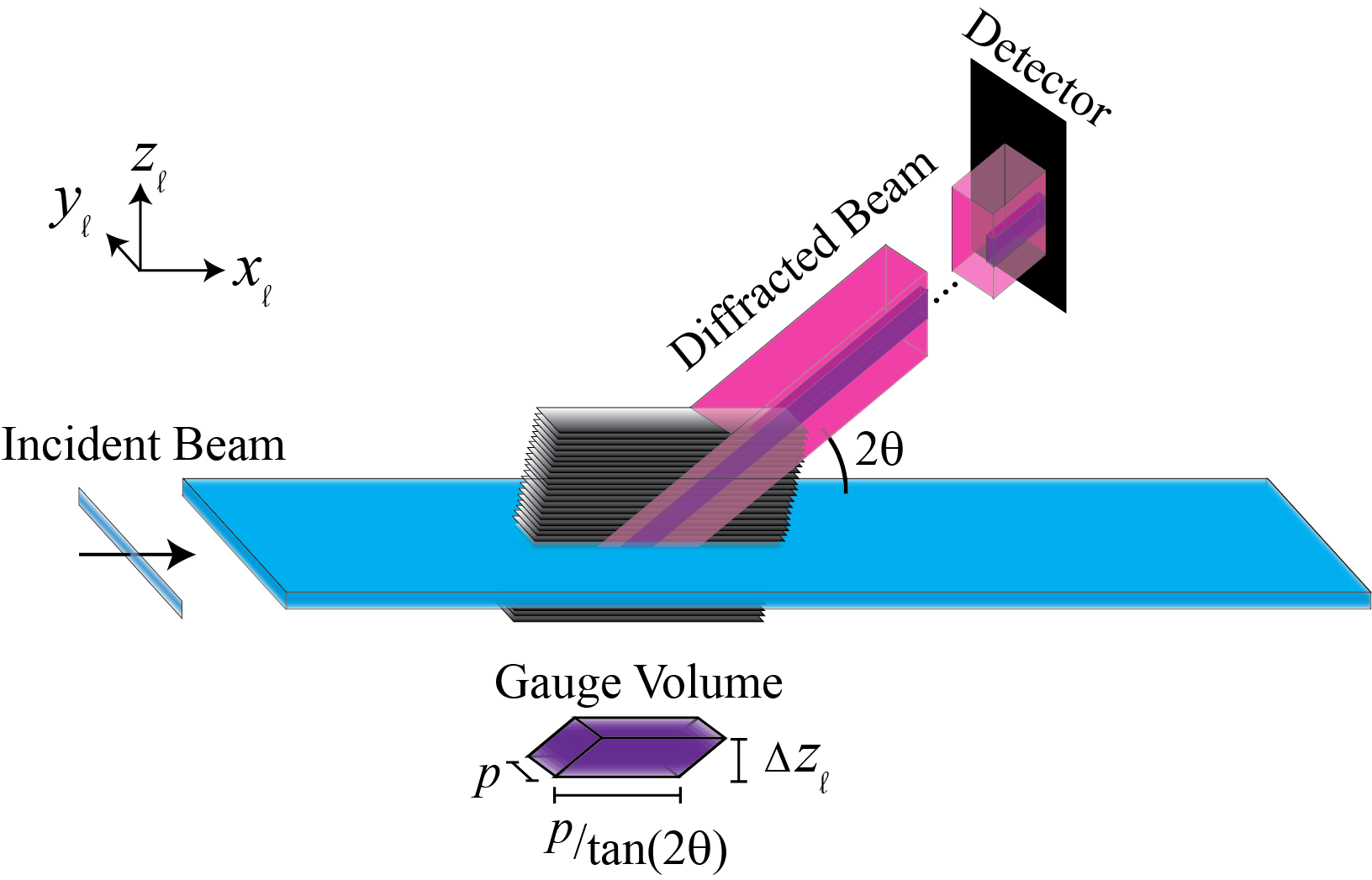}}
    \end{center}
\caption{A schematic illustrating the gauge volume defining the integration in the sample for each pixel on the detector, drawn in the laboratory system. The drawing illustrates that the diffraction angle sets the anisotropic resolution between the $x_{\ell}$ and $y_{\ell}$, as the image is projected along the $2\theta$ angle. The gauge volume is thus defined in the lab system by $p$, $p/\tan(2\theta)$ and $\Delta z_{\ell}$.}
\label{figure-integration}
\end{figure}

For the direct-space resolution function, the angular distribution of the incident beam had not been measured experimentally and only its range was known. We explored several solutions, and will below present result for a Gaussian distribution with a FWHM of $\Delta \zeta_v$.  In the simulations, we initially assume that the detector's PSF is an ideal top-hat function with a width given by the absolute pixel size, such that there is no cross-talk between pixels. This implies that we can associate each detector pixel with a gauge volume in direct space, $V_\ell$, in the sample at the object plane. This corresponds to a ``rounded'' parallelepiped defined by a base-plane spanned by $ (p, p/\tan(2\theta))$ a height of $\Delta z_{\ell}$ and the oblique angle $2\theta$, as illustrated in Fig. \ref{figure-integration}. Inserting the effective gauge volume per pixel becomes $75 \times 204 \times 600$ nm$^3$. 

With these approximations, the forward model for $\phi = 0$ can be constructed based on Eq. \ref{eq-forward4}, with $\rho = 1$  and $\Delta q_{i,1} = 0$ throughout. For relevant offsets in $\phi$ on the ``rocking curve,'' the rotation in \textbf{H} is negligible, but the reciprocal space resolution function is offset. As our simulations assume an idealized detector, we use the results calculated at this point as the final result; future work may extend this analysis to include a final step that blurs the images according to the PSF that is appropriate for the relevant detector.

\subsection{Comparison with Experiment}

\begin{figure}
    \begin{center}
    \resizebox{0.9\columnwidth}{!}{\includegraphics{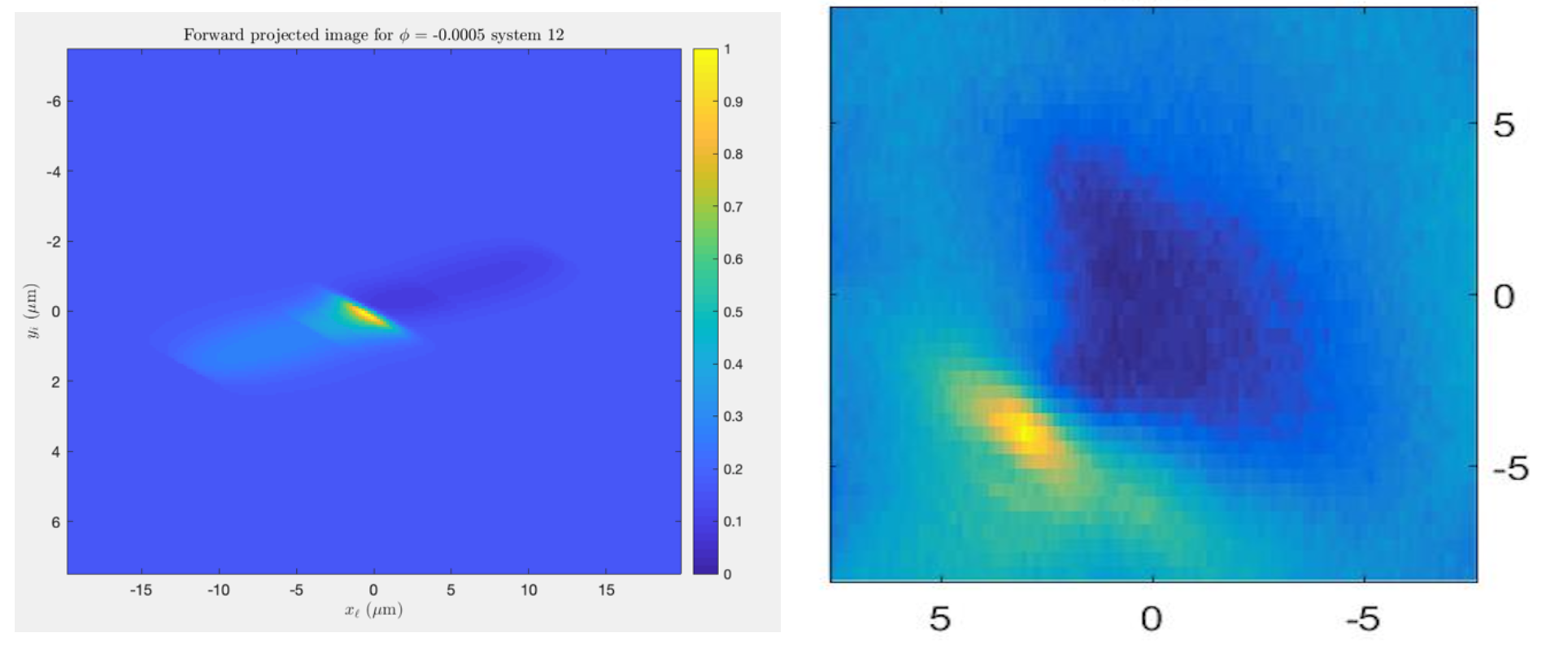}}
    \end{center}
\caption{Forward simulation of a single edge dislocation for $\phi = - 0.5$ mrad (left) and corresponding experimental data (right). The images both refer to the object plane, as plotted in the lab system.}
\label{figure-results}
\end{figure}

The synchrotron experiment we have chosen for demonstration was an actual science study that measured dislocation dynamics within a large single crystal at temperatures near the melt \cite{Dresselhaus2020}, and therefore is not a simplified or idealized setting for comparison. Shown to the right in Fig. \ref{figure-results} is a region of interest (ROI) within one image related to one presumably isolated dislocation. The ROI was defined such that the dislocation is centered in the image. Comparing simulations of each of the 12 primary slip systems found in FCC crystals, we find that simulations related to one of these, match the experiment substantially better than any of the others. Having identified the slip system, we explored several profiles for the divergence of the incident beam to refine our estimates, as they were not measured explicitly. Settling on a Gaussian function, with no further optimization, the forward simulated result is provided to the left in Fig. \ref{figure-results}. 

The correspondence is in our view satisfactory  as it represents a ``real life'' experiment. Notably, thermal drifts may have occurred during the study at elevated temperatures, and the use of a large single crystal implies that dynamical diffraction will be severe when close to the nominal Bragg point. We note that the comparison may be further optimized by adding additional detector noise, and PSFs, as described in Sec.~\ref{section-displacement_ballroom}, however, we found that it was not necessary for the assignment of the dislocation in this system.

%%%%%%%%%%%%%%%%%%%%%%%%%%%%%%%%%%%%%%%%%%%%%%%%%%%%%%%%%%%%%%%%%%%%%%%%%%%%%%

\section{Discussion}

The formalism for a single projection was presented. Procedures for 3D mapping involves tomographic type reconstructions based on scanning $\omega$ \cite{Simons2015} or stacking of 2D slices \cite{Poulsen2020}. Procedures to map orientation and strain variations ranging over angular distances larger than the $NA$ of the objective involves scanning of one or more of the angles $\chi, \phi, \mu$ and $2\theta$ \cite{Poulsen2017}. Forward modelling can then be performed for each projection at a time, as presented in Section \ref{section-forward&scan}.

Forward modelling may be used to optimize the parameters of a material model so that they may effectively match an experiment. As an example, in an iterative scheme positions and directions of dislocations may be optimized by comparing simulated and experimental images. This requires the forward model to be fast; hence the advantage of an lower-dimensional approach, like the one shown in Section \ref{sec-platelike}. Simulations may also be relevant to understand the contrast mechanisms, designing experiments that can suitably measure specific deformations in known crystals, as well as for initial work developing new experimental modalities. 

We note that the formalism and computational approaches outlined in this paper assume that the grain or domain of interest has been indexed, that the reference lattice parameters (and therefore \textbf{B}$_0$) are known and that the optical axis of the objective and trace of the diffracted beam for the $(h,k,\ell)$ diffraction vector for the reference lattice are aligned with high precision.  While it is trivial to fulfill these conditions in forward simulations, additional measurements are required to access the relevant information experimentally, e.g., by global grain mapping using say 3DXRD or DCT prior to DFXM. Notably the reference lattice parameters do not necessarily have to correspond to the values for the strain-free material.  

The work is also conditioned on the use of geometrical optics and on the validity of kinematical diffraction theory. Dynamical diffraction effects are outside the scope of the paper, and merits a thorough investigation both theoretically and experimentally. We remark that such effects are smaller for disordered systems and small grains and reduces with increasing X-ray energy. Moreover, coherent effects may be estimated by experiment or other simulation programs and the effects incorporated in the formalism presented here, e.g. as an additional term in the direct space resolution function.

Parts of this presentation may be relevant for other X-ray diffraction-based methods to facilitate interpretation of micro-structural mapping; opportunities also exist in neutron microscopy \cite{Leemreize2019}. In particular, Bragg Coherent Diffraction Imaging (BCDI) and the related ptychography method also probe one diffraction vector to acquire a high-resolution 3D map the shape and internal displacements in micron-sized isolated particles \cite{Chapman2010,Miao2015,Pfeifer2006, Yau2017,Chamard2015}.  We note that DFXM provides analogous opportunities in bulk samples covering larger length-scales \cite{Poulsen2020}, which could relate more directly to BCDI with this model. Moreover, hybrid methods have recently been proposed for coherency-based mapping of deeply embedded internal volumes, involving the use of an objective: objective based BCDI \cite{Pedersen2020a} and Confocal Bragg Ptychography \cite{Pedersen2020b}. 

DFXM will benefit strongly from the increase in brilliance from the many upgrades of synchrotron sources worldwide currently in progress. Another breakthrough comes from the introducing MLLs as objectives for hard X-rays \cite{Murray2019, Kutsal2019}. As mentioned previously, the numerical apertures of MLLs can be 5-10 times larger than those of the CRLs currently in use. Theoretically this implies spatial resolutions in the range of 10 nm --- an improvement for high-resolution imaging, and a way to map faster (for applications where high $q$-resolution is not needed, e.g. for orientation mapping) . Another advantage is that the reciprocal space resolution function becomes a rectangular top-hat function with a correspondingly 5-10 times larger FWHM. The approximation of a plate-like resolution function will then be fulfilled for most samples. The top-hat resolution function will also simplify sampling procedures both in direct and reciprocal space. 

Finally, we remark that the runtime for the simulations shown in Fig \ref{figure-results} were 30 seconds on a standard stand-alone PC with no optimisation of the code. Hence, with suitable optimisation it appears realistic to simulate images at the same speed as they are acquired at the synchrotron.

%%%%%%%%%%%%%%%%%%%%%%%%%%%%%%%%%%%%%%%%%%%%%%%%%%%%%%%%%%%%%%%%%%%%%%%%%%%%%%

\section{Conclusion}

This work has generalized a 3DXRD framework for analysis of DFXM data, enabling DFXM to interface with 3DXRD experiments via the easy transfer of algorithms and experimental configurations. General-purpose forward modelling procedures have been established by means of analytical expressions and Monte Carlo simulations. Using Monte Carlo methods, we have formulated instrumental resolution functions that are appropriate for a range of different experimental configurations, highlighting how they can be used to facilitate experimental designs.

This work has also connected the DFXM formalism to micro-mechanical models that may be described with continuum mechanics. The insight gained was used to design measurement strategies for mapping individual components of the displacement gradient tensor. We demonstrated the use of this formalism by constructing a forward model of an isolated pure edge dislocation in FCC aluminum. By comparing to experimental data, we demonstrate how this approach may be used to guide the interpretation of DFXM data based on the known parameters of the crystal (e.g. crystal structure, grain orientation, deformations present, etc.). The formalism in this work enables predictive modeling to help design experiments to measure specific types of deformations in materials. 

%%%%%%%%%%%%%%%%%%%%%%%%%%%%%%%%%%%%%%%%%%%%%%%%%%%%%%%%%%%%%%%%%%%%%%%%%%%%%%

\ack{Acknowledgements}

We are grateful to Jon Wright and Hugh Simons for scientific discussions. This work is supported by ERC Advanced Grant nr.~885022 and by the ESS lighthouse on hard materials in 3D, SOLID. We thank ESRF for provision of beamtime and Danscatt for a travel grant. This work was performed in part under the auspices of the U.S.~Department of Energy by Lawrence Livermore National Laboratory under Contract DE-AC52-07NA27344. We also acknowledge the Lawrence Fellowship, which funded contributions from LEDM in this work.

%%%%%%%%%%%%%%%%%%%%%%%%%%%%%%%%%%%%%%%%%%%%%%%%%%%%%%%%%%%%%%%%%%%%%%%%%%%%%%

%\referencelist[references]

\bibliography{}

\end{document}